\documentclass[10pt,letterpaper]{article}
\usepackage[top=0.85in,left=2.75in,footskip=0.75in]{geometry}

\usepackage{changepage}

\usepackage[utf8]{inputenc}

\usepackage{textcomp,marvosym}

\usepackage{fixltx2e}

\usepackage{amsmath,amssymb}

\usepackage{cite}

\usepackage{nameref,hyperref}

\usepackage{microtype}
\DisableLigatures[f]{encoding = *, family = * }

\usepackage{rotating}

\raggedright
\usepackage[aboveskip=1pt,labelfont=bf,labelsep=period,justification=raggedright,singlelinecheck=off]{caption}

\makeatletter
\renewcommand{\@biblabel}[1]{\quad#1.}
\makeatother

\date{}

\usepackage{lastpage,fancyhdr,graphicx}
\usepackage{epstopdf}
\fancyheadoffset[L]{2.25in}
\fancyfootoffset[L]{2.25in}

\begin{document}
\vspace*{0.35in}

\begin{flushleft}
{\Large
\textbf \newline{Protein connectivity in chemotaxis receptor complexes}
}
\newline
\\
Stephan Eismann\textsuperscript{1,2},
Robert G Endres\textsuperscript{2,*},
\\
\bigskip
\bf{1} Department of Physics and Astronomy, University of Heidelberg, Heidelberg, Germany
\\
\bf{2} Department of Life Sciences and Centre for Integrative Systems Biology and Bioinformatics, Imperial College London, London,  UK
\\
\bigskip

%
%





* r.endres@imperial.ac.uk

\end{flushleft}
\section*{Abstract}
The chemotaxis sensory system allows bacteria such as \textit{Escherichia coli} to swim towards nutrients and away from repellents. The underlying pathway is remarkably sensitive in detecting chemical gradients over a wide range of ambient concentrations. Interactions among receptors, which are predominantly clustered at the cell poles, are crucial to this sensitivity. 
Although it has been suggested that the kinase CheA and the adapter protein CheW are integral for receptor connectivity, the exact coupling mechanism remains unclear. 
Here, we present a statistical-mechanics approach  to model the receptor linkage mechanism itself, building on nanodisc and electron cryotomography experiments.
Specifically, we investigate how the sensing behavior of mixed receptor clusters is affected by variations in the expression levels of CheA and CheW at a constant receptor density in the membrane. 
Our model compares favorably with dose-response curves from \textit{in vivo} F\"{o}rster resonance energy transfer (FRET) measurements, demonstrating that the receptor-methylation level has only minor effects on receptor cooperativity. Importantly, our model provides an explanation for the non-intuitive conclusion that the receptor cooperativity decreases with increasing levels of CheA, a core signaling protein associated with the receptors, whereas the receptor cooperativity increases with increasing levels of CheW, a key adapter protein. 
Finally, we propose an evolutionary advantage as explanation for the recently suggested CheW-only linker structures.

\section*{Author Summary}
Receptor clusters of the bacterial chemotaxis sensory system act as antennae to amplify tiny changes in concentrations in the chemical environment of the cell, ultimately steering the cell towards nutrients and away from toxins. 
Despite bacterial chemotaxis being the most widely studied sensory pathway, the exact architecture of the receptor clusters remains speculative, with understanding suffering from a number of paradoxical observations. To address these issues with respect to the protein arrangement in the linkers connecting receptors, we present a statistical-mechanics model that combines insights from electron cryotomography on the linker architecture with results from fluorescence imaging of signaling in living cells. 
Although the signaling data for different expression levels of key molecular components in the linkers seems contradictory at first, our model reconciles these predictions with structural and biochemical data. Finally, we provide an evolutionary explanation for the observation that some of the incorporated linkers do not seem to transmit signals from the receptors.


\section*{Introduction}
\textit{Escherichia coli} cells are able to sense changes in the chemical environment, allowing the bacteria 
to move towards higher concentrations of attractants and lower concentrations of repellents. The chemotaxis system is remarkable for its high sensitivity, wide dynamic range, and precise adaptation while only involving a small number of molecular components \cite{sourjik_spatial_2010,sourjik_receptor_2002,endres_physical_2013}. 
Despite the importance of receptor clustering in accounting for these signaling properties \cite{bray_receptor_1998, mello_allosteric_2005, keymer_chemosensing_2006, shimizu_modular_2010}, there are still unresolved issues with the clusters, in particular with respect to the nature of the coupling mechanism between receptors \cite{endres_polar_2009}.
It has been proposed that receptors assemble into larger arrays via the connection of the kinase CheA and the adapter protein CheW \cite{li_core_2011, bhatnagar_structure_2010}, with potentially complementary effects of membrane-mediated interactions \cite{haselwandter_role_2014}. Unexpectedly, \textit{in vivo} F\"{o}rster  resonance energy transfer (FRET) shows that increasing the expression level of CheA of engineered non-adapting receptors decreases the cooperativity among receptors. In contrast, expressing more CheW increases the cooperativity, albeit in different ranges of expression levels \cite{sourjik_functional_2004}. This raises the question of how these different observations can be reconciled.\\ 

\begin{figure}[!ht]
\includegraphics[width=1.0\textwidth]{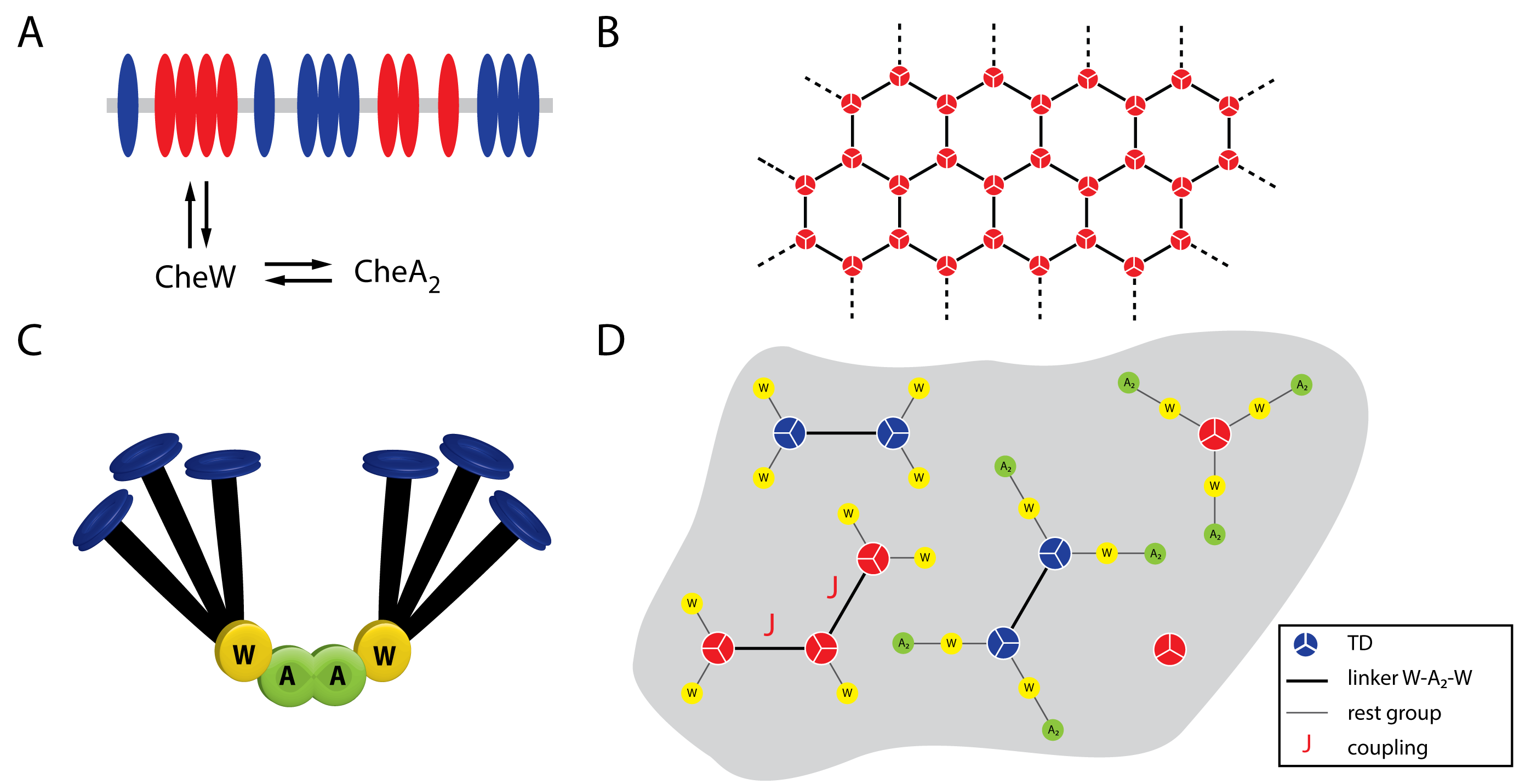} 
\caption{
\label{Fig.Schematics1}
{\bf Schematic of receptor clustering in \textit{E. coli}.}  ({\bf A}) Association and dissociation of adapter protein CheW and kinase CheA\textsubscript{2} with the complexes. CheW interacts directly with the receptors, but the interaction of CheA with the receptors is largely mediated by CheW (see below). ({\bf B}) Electron cryotomography images show that trimers of dimers (TDs) of chemoreceptors cluster at the cell poles in a hexagonal manner \cite{briegel_universal_2009, briegel_bacterial_2012, briegel_new_2014,liu_molecular_2012}.  ({\bf C}) Nanodisc experiments  propose that two TDs are connected by a linker consisting of two CheW monomers and one CheA\textsubscript{2} dimer \cite{li_core_2011}. In reality, the P5 domain of CheA may also contact the trimer \cite{wang_cheareceptor_2012,liu_molecular_2012,piasta_increasing_2014}, although this binding may be an order of magnitude weaker than CheW-trimer binding \cite{li_core_2011, briegel_bacterial_2012}. ({\bf D})  Top view of an ensemble of different sizes of receptor complexes in the cytoplasmic membrane. Active receptors are shown in red, and inactive receptors are shown in blue. Each linker between active TDs contributes a coupling energy $J$.}
\end{figure}

In \textit{E. coli}, there are four types of methyl-accepting chemoreceptors: the high-abundance Tar and Tsr receptors that sense serine and aspartate, respectively, and the low-abundance Trg and Tap receptors \cite{li_cellular_2004, neumann_differences_2010}. In addition, Aer is a chemoreceptor-like sensor of redox potential \cite{bibikov_signal_1997}. The chemoreceptors form homodimers, which assemble into trimers of dimers (TDs) \cite{studdert_crosslinking_2004, boldog_nanodiscs_2006}. On a larger scale, these TDs cluster at cell poles \cite{maddock_polar_1993, sourjik_localization_2000, ames_collaborative_2002}. CheW and CheA, which interact with the cytoplasmic domain of the receptors \cite{piasta_increasing_2014}, are involved in the stabilization of these clusters \cite{kentner_determinants_2006}, which in turn consist of smaller complexes (signaling teams) \cite{keymer_chemosensing_2006, endres_variable_2008, hansen_dynamic-signaling-team_2010}. Signal transduction is triggered by ligand-receptor binding, which leads to a conformational change in the cytoplasmic domains of the receptors  \cite{peach_modeling_2002,vaknin_osmotic_2006, ottemann_piston_1999}. The removal of attractant (or addition of repellent) activates autophosphorylation of the kinase CheA, which is associated with the receptors via the adapter protein CheW (Fig. \ref{Fig.Schematics1}A). The phosphoryl group is then transferred to the response regulator protein CheY, which diffuses through the cytoplasm. CheY-P binds to the flagellar motors to induce clockwise rotation and tumbling of the cell. In contrast, addition of attractant (or removal of repellent) inhibits autophosphorylation of CheA. CheY-P dephosphorylation by phosphatase CheZ leads to counterclockwise rotation and straight swimming \cite{sourjik_spatial_2010}. \\

To avoid saturation of the sensory system, adaptation is implemented via covalent receptor modification. This is achieved through changing the receptor-methylation level by the activities of the methyltransferase CheR and the methylesterase CheB, which antagonistically add and remove, respectively, methyl groups at four or five, depending on the receptor, specific glutamate residues on each receptor monomer \cite{mello_effects_2007}, respectively. Methylation by CheR increases the activity of CheA, i.e., its autophosphorylation rate, thus counteracting the effect of attractant binding. In contrast, CheB activation by phosphorylation by CheA-P decreases CheA activity \cite{sourjik_functional_2004}. Through genetic engineering, the glutamate residues (E) can be replaced by one to four glutamine residues (Q) to mimic increasing receptor-methylation levels in the absence of CheR and CheB \cite{sourjik_receptor_2002}. The \textit{E. coli} chemotaxis pathway is exceptionally well characterized and is thus amenable to modeling at a high quantitative level. \\

To explain the receptor cooperativity, which generates the high sensitivity of the system, the mechanism of receptor-receptor coupling has attracted much interest \cite{shimizu_molecular_2000, endres_polar_2009, greenfield_self-organization_2009, endres_chemotaxis_2007,duke_heightened_1999}. 
Electron cryotomography (EC) images of the TDs in quick-frozen cells led to the idea that TDs form densely packed hexagonal `honeycomb' arrays (Fig. \ref{Fig.Schematics1}B) \cite{briegel_universal_2009, wang_cheareceptor_2012,vu_receptorchew_2012}. 
These and other \textit{in vitro} experiments using nanodiscs and nanoscale plugs to imitate cellular membranes suggest that --CheW--CheA\textsubscript{2}--CheW-- is the structural core unit linking two TDs (see Fig. \ref{Fig.Schematics1}C for a simplified depiction) \cite{li_core_2011}.  
An approach to study the cooperative behavior of the specific receptors inside the cells indirectly is to monitor the signaling activity of CheY-P/CheZ pairs via FRET, with the FRET signal being proportional to the overall CheA activity \cite{sourjik_vivo_2007}. 
An increase in the concentration of CheW was observed to enhance the cooperativity of the FRET response mechanism, whereas, unexpectedly, an increase in CheA concentration led to the opposite effect \cite{sourjik_functional_2004}.
It is well known that multimeric protein complexes can be inhibited by high concentrations of one of their components, similar to the prozone phenomenon in precipitin tests \cite{bray_computer-based_1997}. 
However, it is unclear how the FRET results relate to other experimental observations, including the proposed linker and lattice structures. \\

Here, we use statistical-mechanics modeling within the framework of the Monod-Wyman-Changeux (MWC) model \cite{monod_nature_1965} for cooperative receptor complexes to unify the assumed linker and lattice structures with the seemingly contradictory FRET results.
By implementing the linker structure we initially fit our model of receptor complexes of up to four TDs to FRET data obtained with cells that express only the Tar receptor in different non-adapting modification states. Next, we apply our model to Tar--Tsr--Tap and Tsr--only cells in the non-adapting QEQE modification state, which mimics half-methylated receptors. 
As a result we recover the experimentally observed decrease in cooperativity of the response to serine with increasing CheA concentration, whereas increasing CheW yields the observed enhanced cooperativity. 
Note, other higher order effects of protein overexpression, such as membrane invaginations or interference of CheA/ CheW with clustering, are not included. Our results surmise that the observed opposing trends in cooperativity are based on a critical combination of the correct linker architecture and a constant average complex size. \\

\section*{Model}
\subsection*{Statistical-mechanics model of chemotaxis receptors}
At the heart of our approach lies the MWC model \cite{mello_allosteric_2005, keymer_chemosensing_2006, sourjik_functional_2004}. Chemoreceptors are regarded as two-state systems being either active (on) or inactive (off), with conformation-dependent dissociation constants $K_\text{D}^\text{on}$ and $K_\text{D}^\text{off}$ for a specific ligand. 
As the attractant affinity of inactive receptors is higher than for active receptors ($K_\text{D}^\text{on} \gg K_\text{D}^\text{off}$), the state ratio tips towards inactive receptors with increasing ligand concentration $c$. In contrast, receptor modification $m$ favors the active state in the absence of ligands represented by an energy offset $\Delta \epsilon(m)$. The resulting single-dimer free energies in the active and inactive states are given by  
\begin{equation} \label{eq:fon}
\begin{aligned}
f_\text{on} &= \Delta \epsilon (m) - \ln \left(1 + \frac{c}{K_\text{D}^\text{on}}\right) + \mu \\
f_\text{off} &=  - \ln \left(1 + \frac{c}{K_\text{D}^\text{off}}\right)  + \mu \ , 
\end{aligned}
\end{equation}
with $\mu$ the chemical potential of the receptors in the membrane. All energies are expressed in units of the thermal energy, $k_B T$. 
In our approach, we allow for an ensemble of different complexes with varying complex size $x$ (i.e. number of connected TDs) and partially developed linkers as rest groups $R$ (Fig. \ref{Fig.Schematics1}D). All receptors within a complex are assumed to share the same conformational state because of tight coupling. For simplicity, we consider the  --CheW--CheA\textsubscript{2}--CheW-- linker structure \cite{li_core_2011}, which we incorporate by assigning energies $\mu_\text{W}$ and $\mu_{\text{A}_2}$ for each CheW and CheA\textsubscript{2} molecule integrated in a specific receptor-complex type (see Discussion section for an alternative linker structure). These energies are of the forms
\begin{equation}
\begin{aligned}
\mu_\text{W} &= \ln \left(\frac{\left(K_\text{D}^\text{W} \cdot K_\text{D}^\text{A}\right)^{1/2} }{[W]} \right) \\
\mu_{\text{A}_2} &= \ln \left(\frac{K_\text{D}^\text{A} }{[A]} \right) \ , 
\end{aligned}
\end{equation}
where $[W]$ and $[A]$ indicate monomer concentrations and $K_\text{D}^\text{W}$ and $K_\text{D}^\text{A}$ are dissociation constants for CheW--receptor and CheW--CheA\textsubscript{2} binding, respectively. In particular $[W]$ and $[A]$ are expressed as fractional changes $i$ and $j$ of wild-type expression levels $[W]_0$ and $[A]_0$, respectively:
\begin{equation} 
\begin{aligned}
 \left[W\right](i) &= i \cdot [W]_0 \\
 \left[A\right](j) &= j \cdot [A]_0 \ .
\end{aligned}
\end{equation}
The TD is assumed to be the smallest receptor unit \cite{vaknin_physical_2007, li_core_2011}, and the maximal number of connected TDs is restricted to four, in line with observed Hill coefficients from FRET \cite{sourjik_functional_2004, endres_variable_2008}. (Including larger complex sizes does not alter the model predictions, but increases the computational complexity significantly; see \textit{Materials and Methods}.)  Each dimer can maximally bind to one molecule of CheW, whereas CheA is assumed to not interact with receptor dimers directly. In order to restrain the combinatorial complexity partially developed linkers are only considered in a symmetric manner, i.e. all rest groups are assumed to be identical in a complex. Furthermore, we attribute an attractant energy $J$ to each linker within an active complex, a treatment in line with the previously proposed enhanced coupling among active receptor dimers \cite{hansen_dynamic-signaling-team_2010}, albeit independent of receptor-modification level. 

The resulting free energies for a complex of size $x$ and rest group $R$ are given by (cf. Fig. \ref{Fig.Schematics1}D)
\begin{equation} \label{eq:Fon}
\begin{aligned}
F_\text{on}(x,R) &= 3 x f_\text{on} + (x-1) \left(\mu_{\text{A}_2} + 2 \mu_\text{W} + J \right) + R\left(\mu_{\text{A}_2} , \mu_\text{W} \right) \\
F_\text{off}(x,R)&= 3 x f_\text{off} + (x-1)  \left(\mu_{\text{A}_2} + 2 \mu_\text{W} \right) + R\left(\mu_{\text{A}_2} , \mu_\text{W} \right) \ ,
\end{aligned}
\end{equation}
with $3x$ receptor dimers per complex of size $x$ and $x-1$ linkers. Such a complex has $x+2$ rest groups with $R\left(\mu_{\text{A}_2} , \mu_\text{W} \right)$ given by  
\begin{equation} 
\begin{aligned}
R_1 &= 0\\
R_2\left( \mu_\text{W} \right) &= (x+2) \ \mu_\text{W} \\
R_3\left(\mu_{\text{A}_2} , \mu_\text{W} \right) &= (x+2) \left(\mu_\text{W} + \mu_{\text{A}_2}\right),
\end{aligned}
\end{equation}
for (1) no rest group, (2) a CheW and (3) a CheW and a CheA dimer, respectively. 
The probability $P_S$ for a certain complex type $S\left(x,R\right)$ and its probability $P_S^\text{on}$ of being active follow from standard combinatorial reasoning and the partition function $Z$ \\
\begin{align}
Z & \equiv 1 + \sum_{S} \left(e^{-F_\text{on}(S)}+ e^{-F_\text{off}(S)}\right) \label{eq:Z} \\
P_S & = \frac{e^{-F_\text{on}(S)}+e^{-F_\text{off}(S)}}{Z} \label{eq:Ps}  \\
P_S^\text{on}  & = \left(1+e^{F_\text{on}(S)-F_\text{off}(S)}\right)^{-1} \ ,\label{eq:Pson}
\end{align}
where the number 1 in the partition function $Z$ reflects the possibility of an empty membrane site. \\

Assuming the FRET signal to report the number $n_{A_2}(S)$ of CheA\textsubscript{2} dimers within an active complex, we define the receptor activity as
\begin{equation} \label{eq:A}
A = \sum_S P\left(S , \text{on}\right) \cdot n_{A_2}(S) = \sum_S P_S \cdot P_S^\text{on} \cdot n_{A_2}(S) \ .
\end{equation}
In contrast, the classical MWC model for coupled receptors describes the response of a single complex of $N$ TDs to a change in ligand concentration. Without incorporating the receptor coupling explicitly, the corresponding activity $\check{A}$ reads \cite{endres_variable_2008}
\begin{equation} \label{eq:Acheck}
\check{A} = \left(1 + \exp \left[N \left\{ \Delta \epsilon(m) + \log\left( \frac{1+ c/ K_\text{D}^\text{off}}{1 + c/K_\text{D}^\text{on}}\right)\right\}\right]\right)^{-1} \ . 
\end{equation}
In the past, the Hill coefficient $n_\text{H}$ and complex size $N$ have broadly been treated as equivalent to quantify the cooperative behavior of receptor complexes, and in \cite{endres_variable_2008}, an increase in $N$ with receptor-modification level was equated with an increase in receptor cooperativity. However, both quantities are not necessarily the same as approximating Eq.~\ref{eq:Acheck} by a Hill function with  $n_\text{H}=N$ requires $c \ll K_\text{D}^\text{on}$ \cite{keymer_chemosensing_2006}.  
We found that, in the classical MWC model, the response of differently modified Tar receptors to MeAsp, a non-metabolizable analog of aspartate, can also be described with a fixed $N$ for all modification levels. This treatment results in a similar quality of fit when relating the reduced number of parameters to the new $\chi^2$ goodness-of-fit value (see \nameref{S1_Fig.}.).
As our model incorporates an ensemble of complexes of varying sizes, the finding of a constant complex size $N$ in the classical MWC model is naturally generalized by a constant average complex size $\left\langle N \right\rangle$ with respect to ligand concentration and receptor-modification state. 
The average complex size, which we term receptor density $\rho$, is given by
\begin{equation} \label{eq:rho}
\rho = \sum_S 3 \cdot x \cdot P_S = 3 \left\langle x \right\rangle \equiv \mathrm{constant} \ ,
\end{equation}
with $x$ being the number of dimers of a given complex type $S$. 
The chemical potential $\mu$ in Eq.~\ref{eq:fon} is adjusted throughout the simulation to fulfill this condition, reflecting anticipated regulation of the receptor-expression level by the cell. Biologically, a constant receptor density can be achieved by random receptor insertion into a growing membrane at constant rate \cite{greenfield_self-organization_2009}. Since wild-type cells express and insert receptors in the QEQE modification state \cite{sourjik_receptor_2002}, we do not expect a modification-dependent insertion rate. Although allowing for a modification-dependent $\rho$ would increase the quality of fit because of an increased number of fitting parameters, our minimal model with constant $\rho$ can describe the data very well.\\

\section*{Results}
\subsection*{Receptor-modification level may not determine cooperative behavior of complexes}
\begin{figure}[!ht]
\includegraphics[width=1.0\textwidth]{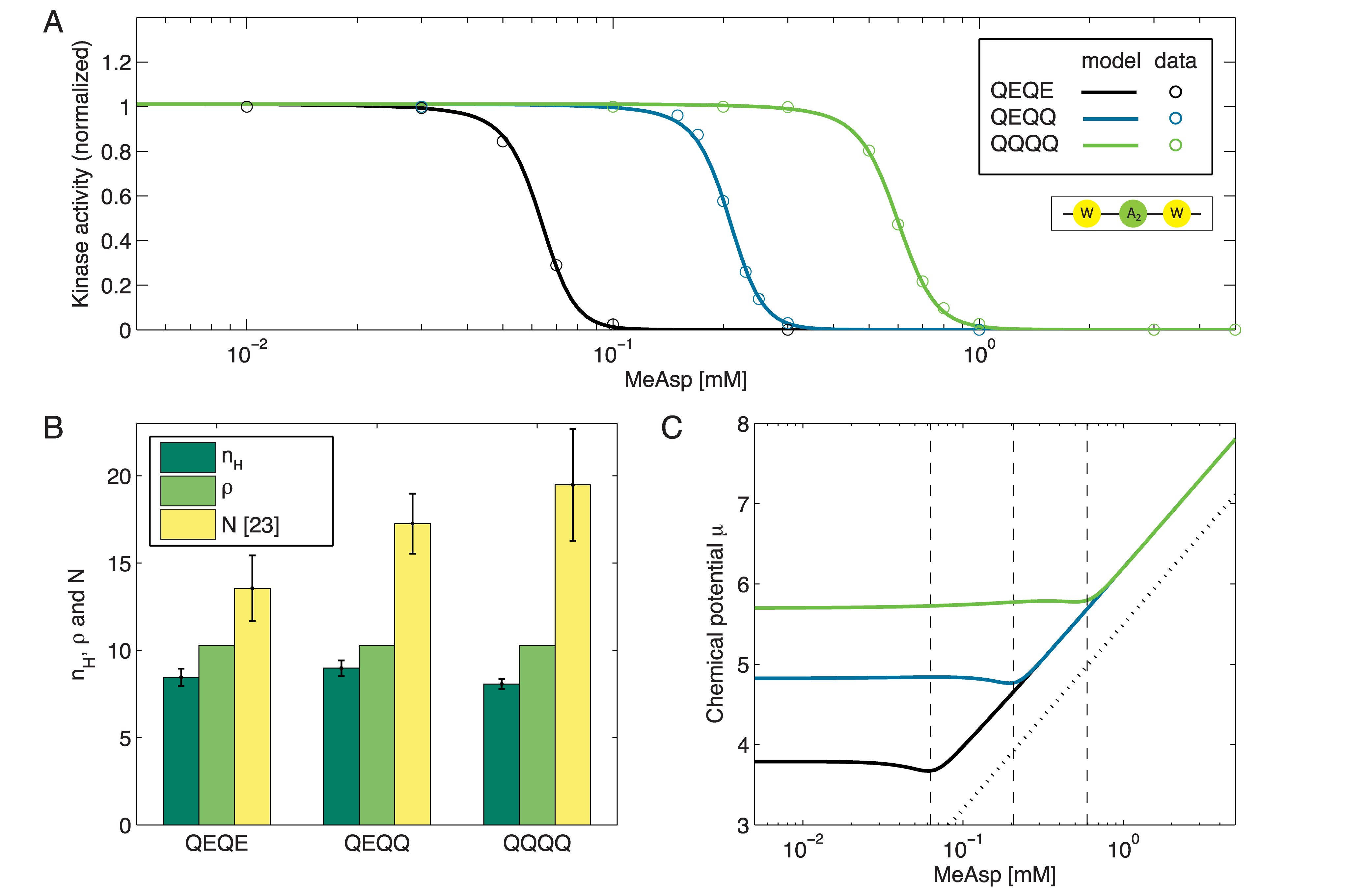}
\caption{
{\bf Kinase activity for different Tar-modification levels.}  ({\bf A}) Kinase activity for Tar receptors in QEQE (black), QEQQ (blue) and QQQQ (green) modification states as a function of MeAsp concentration. The curves are normalized with respect to QQQQ activity at concentration $c=10^{-4}$mM. ({\bf B}) All model curves share the same receptor density $\rho$ (light green), which reflects the average complex size, depicted next to the corresponding Hill coefficients $n_\text{H}$ (dark green). Parameter $N$ (yellow) of the classical MWC model (taken from \cite{endres_variable_2008}) is shown for comparison. ({\bf C}) The chemical potential $\mu$, shown as a function of ligand concentration for the three modification levels, is adjusted throughout the simulation to ensure constant $\rho$ at all concentrations. Color coding is the same as in panel A. The vertical dashed lines indicate half-maximum concentrations $c_\text{H}$ from the corresponding Hill fits. While for $c < c_\text{H}$ $\mu$ is approximately constant, the curves follow a logarithmic function in the regime of $c > c_\text{H}$. The dotted blue line shows $f(c)=\ln{c} + 5.5$ for comparison. Model parameters: $\Delta\epsilon\left(QEQE\right)=-1.12$, $\Delta\epsilon\left(QEQQ\right)=-2.16$, $\Delta\epsilon\left(QQQQ\right)=-3.03$, $K_\text{D,Tar}^\text{on}=2.18$, $K_\text{D,Tar}^\text{off}=0.001$, $\rho=10.30$, $\mu_{\text{W}}^0=-0.67$, $\mu_{\text{A}_2}^0=-1.68$, and $J=-3.81$. Values for $\mu_{\text{W}}^0$, $\mu_{\text{A}_2}^0$, and $J$ are shared with curves shown in Figs.  \ref{Fig.ChangeCheW} and  \ref{Fig.ChangeCheA}. The superscript 0 indicates wild-type expression levels for CheA/CheW.}
\label{Fig.Tar-only}
\end{figure}

In order to test our model, we firstly applied it to FRET data of Tar-only receptors in different non-adapting receptor-modification states from Ref. \cite{endres_variable_2008} i.e. Tar\{QEQE\}, Tar\{QEQQ\} and Tar\{QQQQ\}.  
The dose-response curves of the chemoreceptors match closely the statistical-mechanics model with fixed receptor density, and hence fixed average complex size (Fig. \ref{Fig.Tar-only}A).  
Figure \ref{Fig.Tar-only}B displays the fitted receptor density $\rho$ next to the Hill coefficients $n_\text{H}$ of the experimental curves (see \textit{Materials and Methods}) and the complex size $N$ of the classical MWC model, taken from \cite{endres_variable_2008}. Although the classical MWC model predicts a rise in complex size with modification level \cite{endres_variable_2008}, including its implementation based on a dynamic Ising model \cite{hansen_dynamic-signaling-team_2010}, this is not true for the Hill coefficients (see also \nameref{S1_Fig.}.). This finding shows that receptor modification is not the main determinant of receptor cooperativity.

In our model, the chemical potential $\mu$ can be regarded as the cost function for the cell to provide a constant complex size in the membrane. By definition, the chemical potential $\mu \equiv \partial F/\partial N$ reflects the amount of energy required for adding a particle to a system with free energy $F$. 
Although the value of the parameter $\mu$, introduced to ensure constant receptor density $\rho$, is gained by solving a highly nonlinear equation, its behavior with respect to ligand concentration is very homogeneous and characterized by two regimes, as shown in Fig. \ref{Fig.Tar-only}C. 
While this cost is approximately constant for $c < c_\text{H}$, with $c_\text{H}$ being the half-maximum concentration obtained from Hill fits, the cost necessary to maintain a constant density increases rapidly for ligand concentrations beyond $c_\text{H}$. In this second regime, the curves for all modification levels $m$ are of the form $f(c) = f_0 + \ln c$, which is the functional description of an ideal chemical potential.  Although the slope in the second regime is the same for all values of $m$, the different offsets $f_0\left(m\right)$ reflect the modification-dependent energy $\Delta \epsilon (m)$. Note, if we were instead to keep $\mu$ constant (and not $\rho$), then bumps would appear in the dose-response curves as a result of the receptor density increasing with ligand concentration (see \nameref{S2_Fig.}.).\\

In summary, our model is capable of quantitatively describing dose-response curves from \textit{in vivo} FRET, in particular the receptor-receptor cooperativity. 
Although in spirit similar to other recent statistical-mechanics models, most noticeably by \textit{Hansen et al.} \cite{hansen_dynamic-signaling-team_2010} and \textit{Lan et al.} \cite{lan_adapt_2011}, only our model addresses the protein connectivity in receptor complexes. 

\subsection*{Receptor density governs cooperative behavior of complexes}
 
\begin{figure}[!ht]
\includegraphics[width=1.0\textwidth]{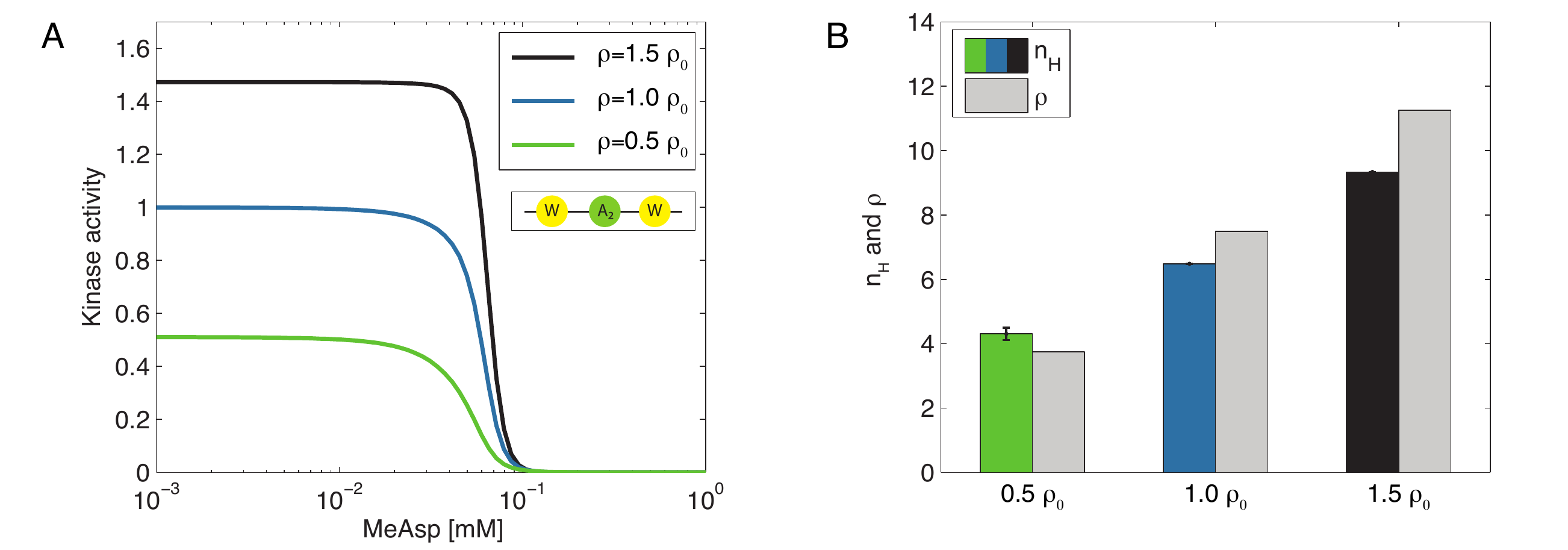}
\caption{
{\bf Cooperativity increases with receptor density.}  ({\bf A}) Model curves based on previously fitted parameters of Tar--only cells in QEQE modification state (Fig. \ref{Fig.Tar-only}) for different receptor densities $\rho=1.5\cdot \rho_0$ (black), $\rho=\rho_0$ (blue) and $\rho=0.5\cdot \rho_0$ (green) with the wild-type receptor density $\rho_0=7.5$.  ({\bf B}) Hill coefficients $n_H$ derived by fitting to the model outcome and corresponding receptor densities $\rho$. \\
}
\label{Fig.Rho}
\end{figure}

While the receptor density $\rho$ is assumed to be constant on a short time scale, the rate of receptor expression and insertion into the membrane can be regulated by the cell on a longer time scale.
Hence, as a further test of our statistical-mechanics model, we investigated how a change in receptor density $\rho$ affects CheA activity at wild-type expression levels for CheA and CheW. 
Figure \ref{Fig.Rho}A shows modeled dose-response curves for different $\rho$ values of $1.5\cdot \rho_0$ , $\rho_0$  and $0.5\cdot \rho_0$ with $\rho_0=7.5$ the wild-type receptor density and otherwise using the same parameter set as in Fig. \ref{Fig.Tar-only}. An increase in receptor density is directly associated with an enhanced signal amplitude because more CheA molecules are incorporated into the complexes.
Figure \ref{Fig.Rho}B reflects the associated trend in cooperativity by comparing density $\rho$ and Hill coefficient $n_H$. In qualitative agreement with experimental observations \cite{sourjik_functional_2004} and in line with previous modeling \cite{keymer_chemosensing_2006}, larger complex sizes lead to higher sensitivities and hence steeper dose-response curves given a certain receptor-modification state. Since the expression level of receptors (and other chemotaxis proteins) is highest under nutrient-poor conditions, the resulting increase in receptor density and cooperativity leads to enhanced sensitivity when it is most crucial for cell survival \cite{khursigara_lateral_2011}. 

\subsection*{Increasing the CheW level increases receptor cooperativity}
To gain insight into the role of CheA and CheW in forming receptor complexes, we varied the expression levels [A] and [W] to study the effect on receptor activity. According to the experimental observations in \cite{sourjik_functional_2004}, we set the CheW concentrations to 0.7, 0.1 and 0.01 and the CheA concentrations to 8, 0.3 and 0.25 times the wild-type values $[W]_0$ and $[A]_0$, respectively. This allowed us to make the comparison with experimental dose-response curves from FRET of Tsr--only cells (for varying CheW) and Tar--Tsr--Tap cells (for varying CheA), both in the non-adapting QEQE modification state.
To keep the overall number of parameters small, the data for changes in [A] and [W] was fitted with the same parameter set ($\Delta\epsilon$, $\rho$, $K_\text{D,Tsr}^\text{on}$, $K_\text{D,Tsr}^\text{off}$, $J$, $\mu_{\text{W}}^0$ and $\mu_{\text{A}_2}^0$). 
Multiplication of the calculated activities with scaling parameters $s_\text{A}$ and $s_\text{W}$, respectively allows for comparison with the FRET signal amplitudes. Subsequently, a Hill function was fitted to the model curves and the model Hill parameters were compared with the experimental values. Note that our minimal model does not account for alternative forms of signaling disruption upon over- or underexpression of CheA/CheW, such as zipper-like invaginations of the cell membrane \cite{zhang_direct_2007} or interference with trimer formation \cite{studdert_crosslinking_2004}. 
 
\begin{figure}[!ht]
\includegraphics[width=1.0\textwidth]{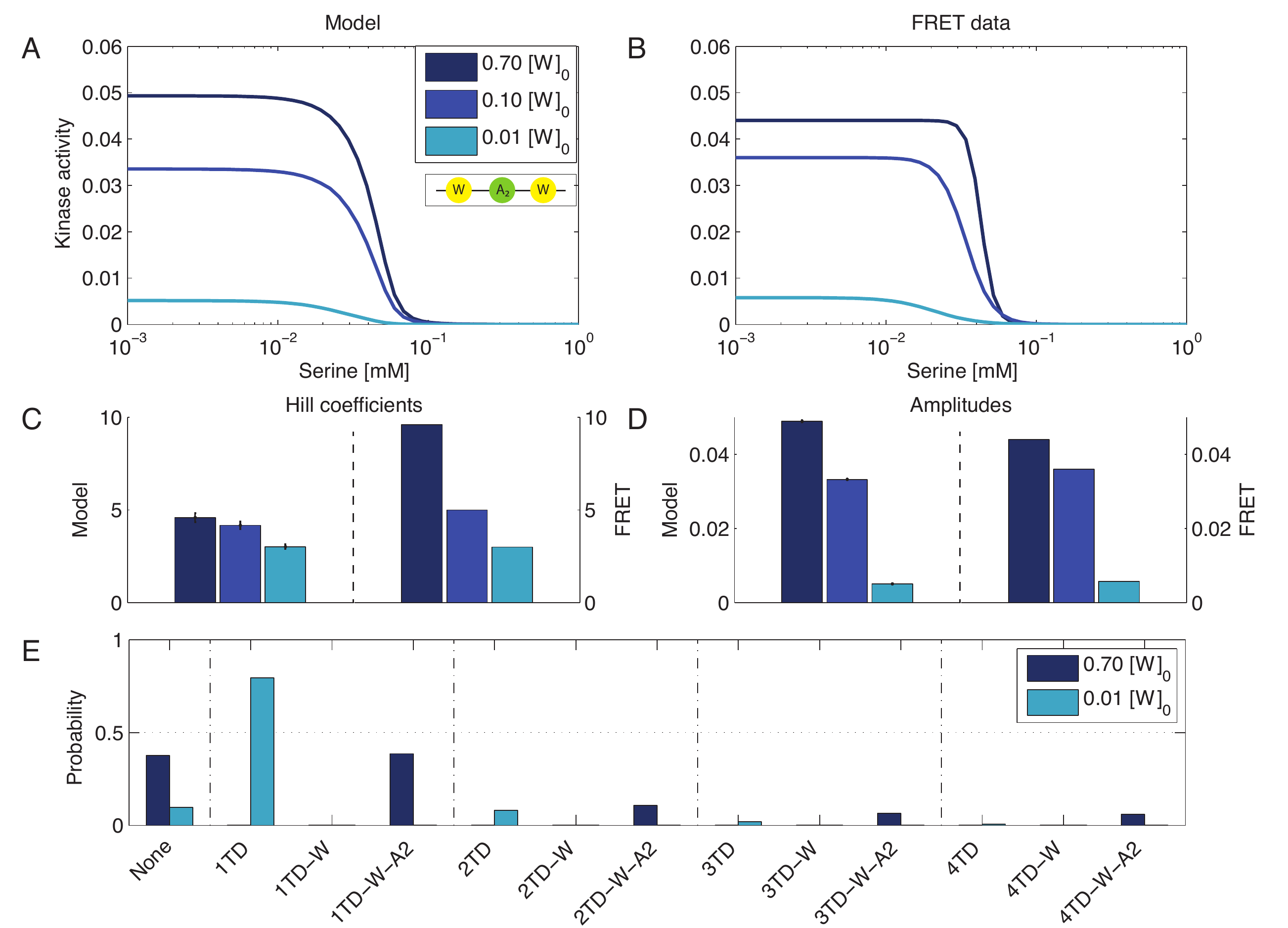}
\caption{
{\bf Cooperativity increases with the expression level of CheW.}  ({\bf A,B}) Model fit (A) and FRET data \cite{sourjik_functional_2004} (B) for different expression levels of CheW at 0.7 (dark blue), 0.1 (blue) and 0.01 (azure) times the native level of $[W]_0$.  ({\bf C}) Hill curves were fitted to the model outcome to allow for comparison with experimental results. Hill coefficient pairs (model/experiment) in order of increasing [W] are (3.0/3.0), (4.2/5.0) and (4.6/9.6). ({\bf D}) The corresponding Hill amplitudes in order of increasing [W] are (0.005/0.006), (0.033/0.036) and (0.049/0.044). ({\bf E}) Distribution of complex types present at half-maximum concentration for 0.7 (dark blue) and 0.01 (azure) times the native concentration $[W]_0$. Model parameters: $\Delta\epsilon\left(QEQE\right)=-2.42$, $K_\text{D,Tsr}^\text{on}=2.18$, $K_\text{D,Tsr}^\text{off}=0.002$, $\rho=3.13$, $\mu_{\text{W}}^0=-0.67$,  $\mu_{\text{A}_2}^0=-1.68$, and $J=-3.81$. Parameters are shared with model for variation in [A] (Fig. \ref{Fig.ChangeCheA}). 
}
\label{Fig.ChangeCheW}
\end{figure}

Figures  \ref{Fig.ChangeCheW}A,B show the model data next to the experimentally determined Hill curves for variations in [W]. 
Enhanced CheW expression results in raised activity amplitudes and Hill coefficients (Fig. \ref{Fig.ChangeCheW}C,D). Although the $n_\text{H}$ values from the model change significantly with expression level [W] at a $95\%$ confidence level, which is in qualitative agreement with the experimental data, especially with respect to the highest CheW expression level, the change in $n_\text{H}$ is less pronounced for the model than the experimental data. The positive correlation between kinase activity and amount of available CheW becomes evident in the distribution of complex species at half-maximum concentration (Fig. \ref{Fig.ChangeCheW}E). Whereas low levels of [W] favor independent, single TDs, larger complexes are more likely to form for larger [W]. As the probability for an empty membrane site also increases, the receptor density remains constant. 

\subsection*{Increasing the CheA level decreases receptor cooperativity}
\begin{figure}[!ht]
\includegraphics[width=1.0\textwidth]{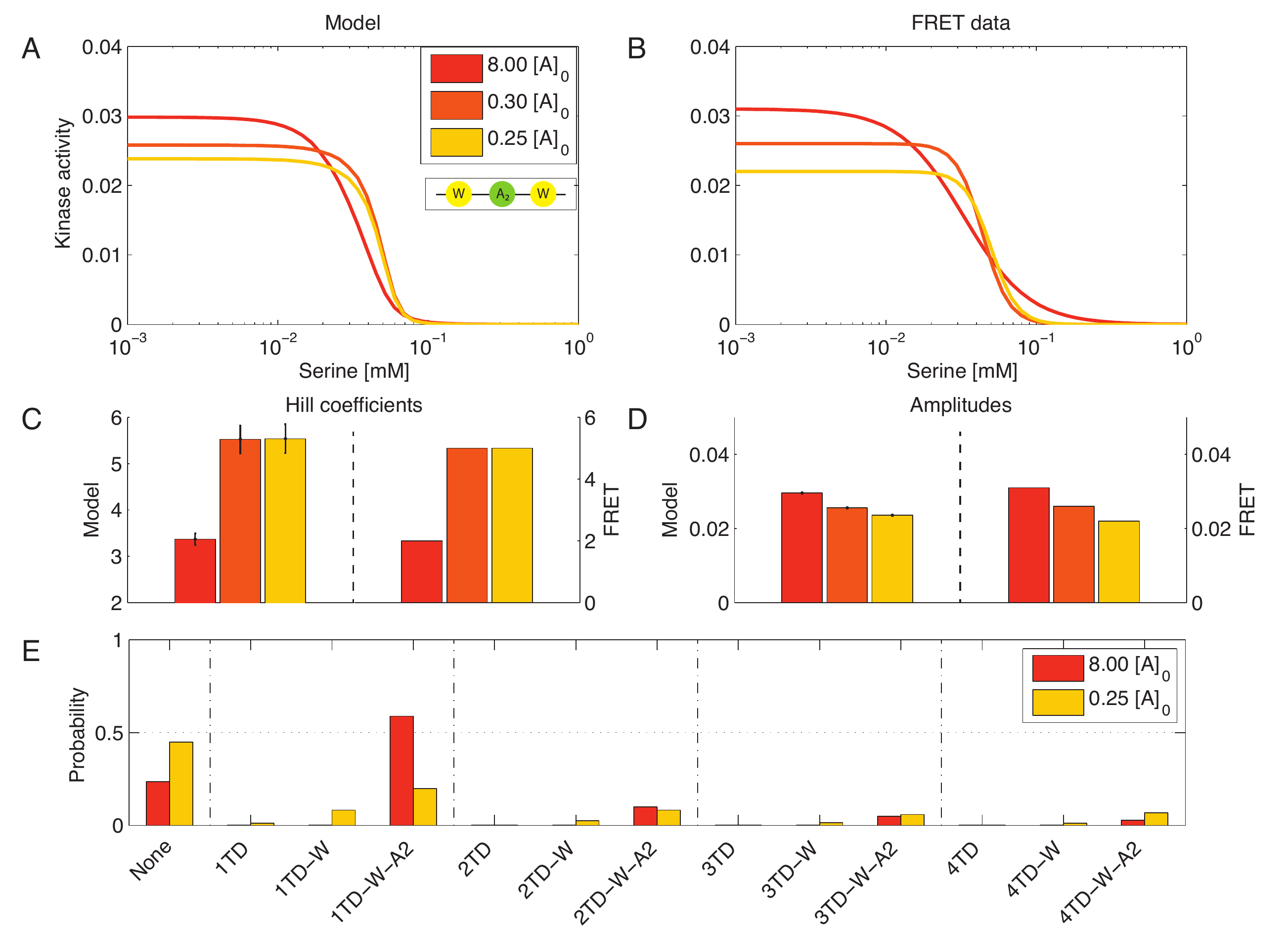}
\caption{
{\bf Cooperativity decreases with the expression level of CheA.}  ({\bf A,B}) Model fit (A) and FRET data \cite{sourjik_functional_2004} (B) for different expression levels of CheA at 8 (red), 0.3 (orange) and 0.25 (yellow) times native level $[A]_0$.  ({\bf C}) Hill curves were fitted to the model outcome to allow for comparison with experimental results. Hill coefficient pairs (model/experiment) in order of increasing [A] are (5.5/5.0), (5.5/5.0) and (3.4/2.0). ({\bf D}) The corresponding Hill amplitudes in order of increasing [A] are (0.023/0.022), (0.026/0.026) and (0.030/0.031). ({\bf E}) Distribution of complex types present at half-maximum ligand concentration for 8 (red) and 0.25 (yellow) times native concentration $[A]_0$. Model parameters: $\Delta\epsilon\left(QEQE\right)=-2.42$, $K_\text{D,Tsr}^\text{on}=2.18$, $K_\text{D,Tsr}^\text{off}=0.002$, $\rho=3.13$, $\mu_{\text{W}}^0=-0.67$,  $\mu_{\text{A}_2}^0=-1.68$, and $J=-3.81$. Parameters are shared with model for variation in [W] (Fig. \ref{Fig.ChangeCheW}).
}
\label{Fig.ChangeCheA}
\end{figure}

Changing [A] in our model has the opposite effect on the Hill coefficient as changing [W]. This result is in line with experimental data (Fig. \ref{Fig.ChangeCheA}A,B,C). The activity amplitude reflecting the amount of active CheA molecules benefits from higher CheA levels, as one would expect (Fig. \ref{Fig.ChangeCheA}D). In contrast, Hill coefficients are higher for smaller [A], recovering the naively unexpected experimental observations (Fig. \ref{Fig.ChangeCheA}C). Looking at the distribution of complexes at half-maximum ligand concentration (Fig. \ref{Fig.ChangeCheA}E), we note that although high CheA concentrations favor rest groups including CheA, complex sizes of 3 and 4 TDs are more likely at lower concentrations of CheA. 

\begin{figure}[!ht]
\includegraphics[width=1.0\textwidth]{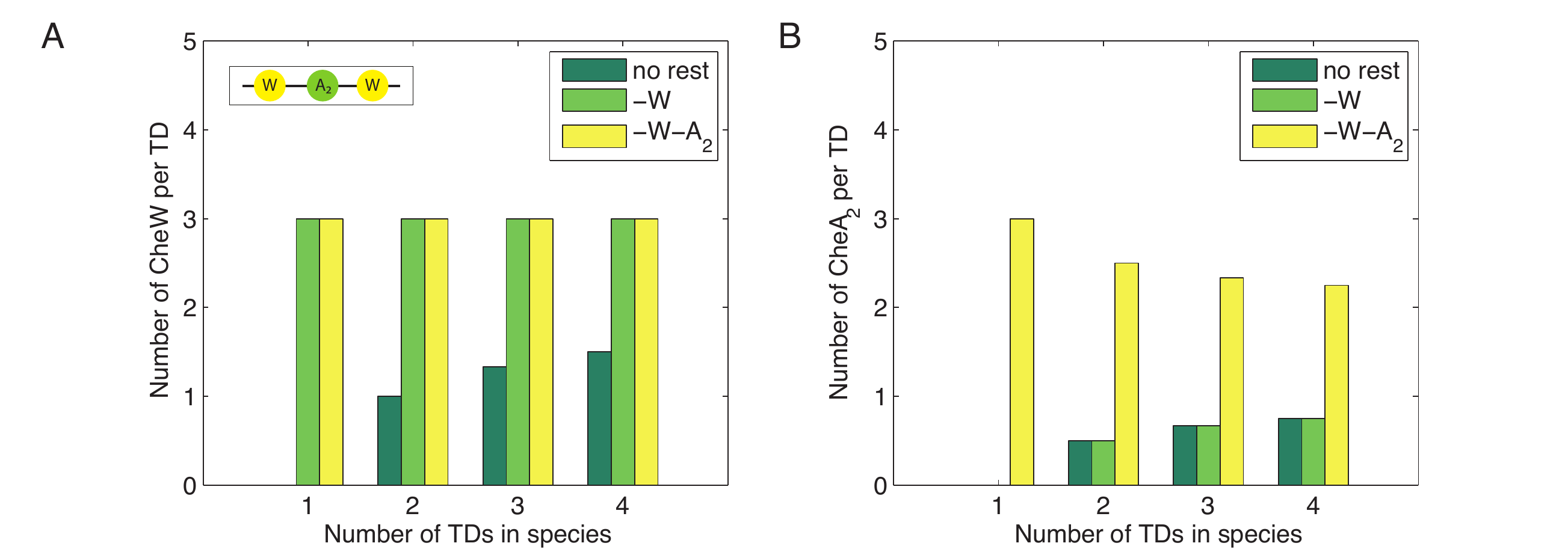}
\caption{
{\bf CheW and CheA counts per TD show different trends.}  ({\bf A}) For complexes with rest groups, the number of CheW molecules per TD is independent of complex size. For species with no rest groups, this ratio increases with the number of TDs in the species. Raising the expression level of CheW results in the formation of larger complex sizes. ({\bf B}) The number of CheA dimers per TD shows two opposing trends with respect to complex size. As the ratio is highest for single TDs with CheA\textsubscript{2}-including rest groups, raising the expression level of CheA results in smaller complex sizes but an increased number of CheA molecules contributing to the signal amplitude. 
}
\label{Fig.Counting}
\end{figure}

The opposing trends in $n_\text{H}$ concerning variations in [A] and [W] are a direct result of the linker stoichiometry and fixed average complex size. 
For complexes with rest groups, the ratio of CheW molecules per TD is independent of the complex size (Fig. \ref{Fig.Counting}A). However, for species without rest groups, this ratio increases with the number of coupled TDs. As a result, an enhancement in [W] yields larger complexes that directly incorporate more CheA molecules. Furthermore, empty sites ensure a constant receptor density even when expression levels of CheW and CheA are extremely low. In this case, the receptor density still remains constant as empty sites can be occupied by individual TDs. This requires a dilute membrane, i.e., a receptor density not much larger than $\left\langle \rho \right\rangle = 3 \left\langle x \right \rangle = 9$ (see Fig. \ref{Fig.Tar-only}B).

In contrast, the corresponding ratio of CheA dimers per TD is highest for single TDs with full rest groups and decreases with increasing complex size (Fig. \ref{Fig.Counting}B). The CheA molecules within the rest groups contribute to the FRET amplitude but not to the receptor cooperativity. An accompanying rise in the number of occupied membrane sites ensures a constant receptor density.

\subsection*{Electron cryotomography suggests the existence of CheW-only linkers}
\begin{figure}[!ht]
\includegraphics[width=1.0\textwidth]{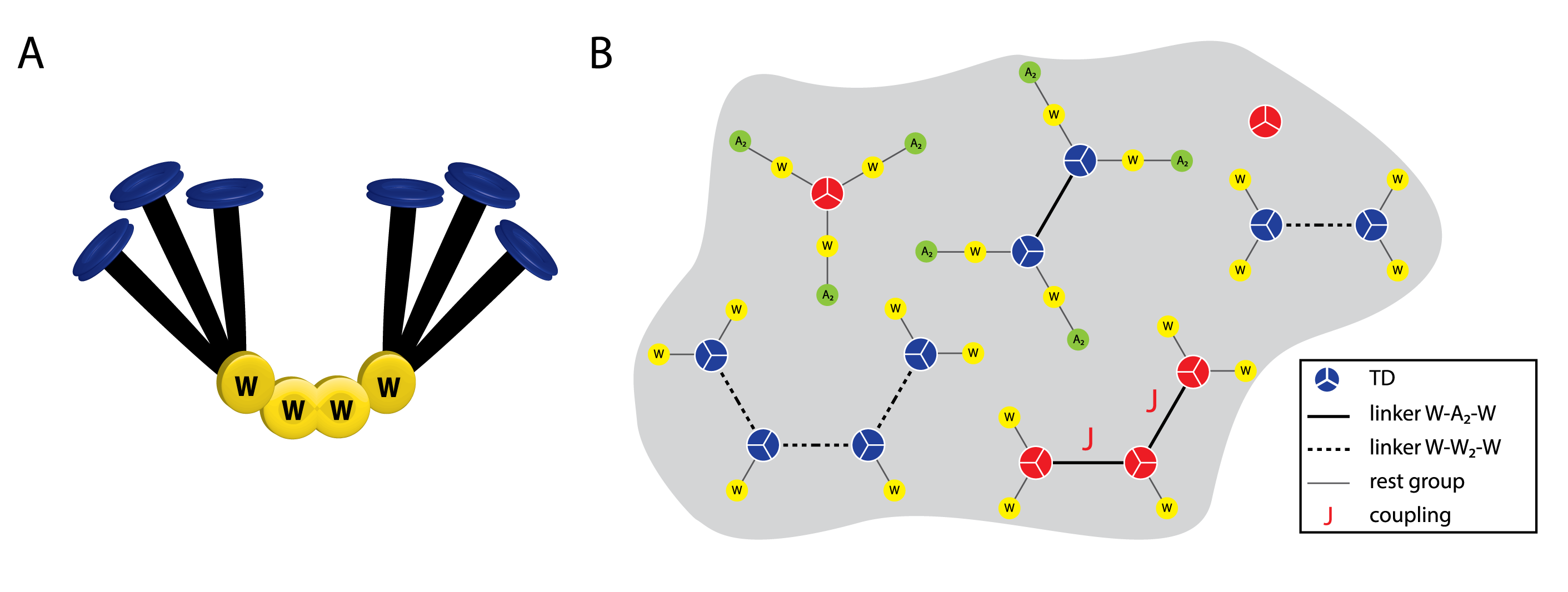}
\caption{
{\bf CheW-only linkers.}  ({\bf A}) In agreement with recent electron cryotomography experiments \cite{liu_molecular_2012, briegel_new_2014}, we allow for additional CheW-only linkers (yellow) connecting two TDs (black and blue). 
  ({\bf B}) Exemplary ensemble of complexes in the cytoplasmic membrane. The two linkers are represented by solid (--CheW--CheA\textsubscript{2}--CheW--) and dashed (--CheW--CheW\textsubscript{2}--CheW--) lines. Active and inactive TDs are shown in red and blue, respectively. Each linker between active TDs contributes an additional coupling energy $J$. 
}
\label{Fig.SchematicNew}
\end{figure}

Our model qualitatively reproduces the experimental results obtained when the expression levels of CheW and CheA were changed. 
However, there are quantitative differences, especially with respect to the change in cooperativity as a function of the expression level of CheW. This change is less pronounced in the model than in the experiment. Recent findings from electron cryotomography may shed light on the reasons for these discrepancies.
Although both studies stressed the importance of one dimeric CheA and two CheWs as the minimal unit needed for kinase activation, \textit{Briegel et al.}\cite{briegel_new_2014} and \textit{Liu et al.}\cite{liu_molecular_2012} proposed additional CheW-only linkers, underlining the role of CheW in the cooperative behavior of TDs. Such structures could explain how increased levels of CheW contribute to the cooperativity of TDs.
In order to quantify this effect, we allowed for additional CheW-only linkers in our model (Fig. \ref{Fig.SchematicNew}).
The dimeric appearance of CheW in the linker is accounted for by a new parameter $\mu_{\text{W}_2}$; we keep the previously introduced rest groups for simplicity. 

\begin{figure}[!ht]
\includegraphics[width=1.0\textwidth]{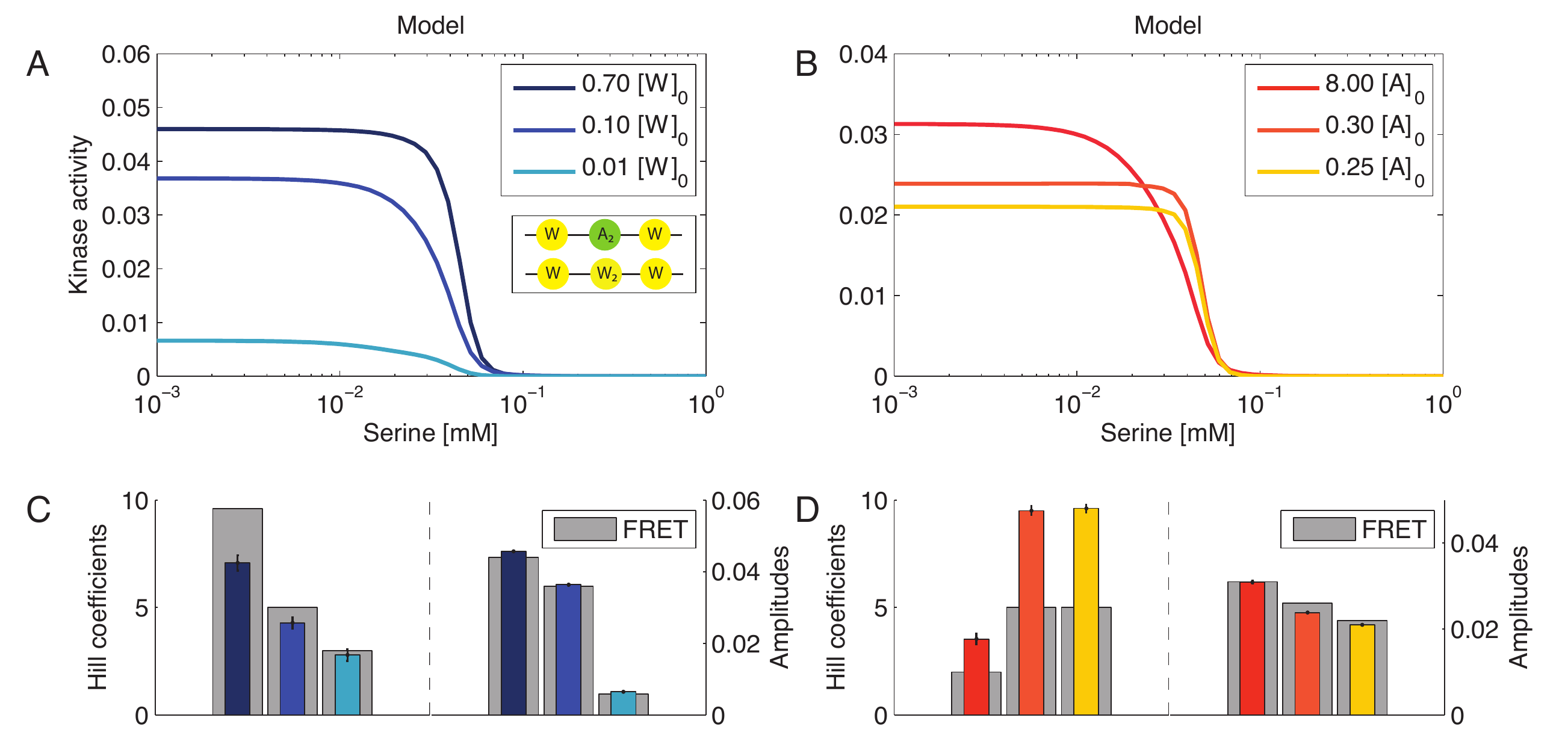}
\caption{
{\bf CheW-only linkers can explain the large enhancement in cooperativity with increasing [W].}  ({\bf A,B}) Modeled dose-response curves for different levels of expression of CheW (A) and CheA (B) as multiples of wild-type levels $[W]_0$ and $[A]_0$, respectively. ({\bf C,D}) Experimental results in gray are superimposed with parameters inferred from Hill curves fitted to the model outcome. Parameters for variation in [W] and [A] are shown in panels (C) and (D), respectively. 
Model parameters: $\Delta\epsilon\left(QEQE\right)=-1.79$, $K_\text{D,Tsr}^\text{on}=3.53$, $K_\text{D,Tsr}^\text{off}=0.003$, $\rho_{\text{W}}=3.52$, $\rho_{\text{A}}=4.45$, $\mu_{\text{W}}^0=-0.83$,  $\mu_{\text{A}_2}^0=-1.65$, $\mu_{\text{W}_2}^0=-5.02$, and $J=-4.07$. The data for variations in CheA and CheW levels was fitted with receptor densities $\rho_W$ and $\rho_A$, respectively.
}
\label{Fig.ResultsNew}
\end{figure}

Figure \ref{Fig.ResultsNew} shows the results for varying expression levels of CheW and CheA.  
The dose-response curves of the new model exhibit the same trends in Hill coefficient and amplitude for variation in [W] (Fig. \ref{Fig.ResultsNew}A) and [A] (Fig. \ref{Fig.ResultsNew}B) as before, in agreement with experimental results (see also \nameref{S3_Fig.}.). 
However, the difference in behavior is manifested in the comparison panels below. The previously obtained minor changes in receptor cooperativity as a function of [W] are now much more pronounced (Fig. \ref{Fig.ResultsNew}C), although the modeled Hill coefficients for [A] variation are larger than the experimental ones (Fig. \ref{Fig.ResultsNew}D). 
The excess CheW leads to formation of CheW-only linkers and hence larger complex sizes when the amount of available CheA is held constant. 

\begin{figure}[!ht]
\includegraphics[width=1.0\textwidth]{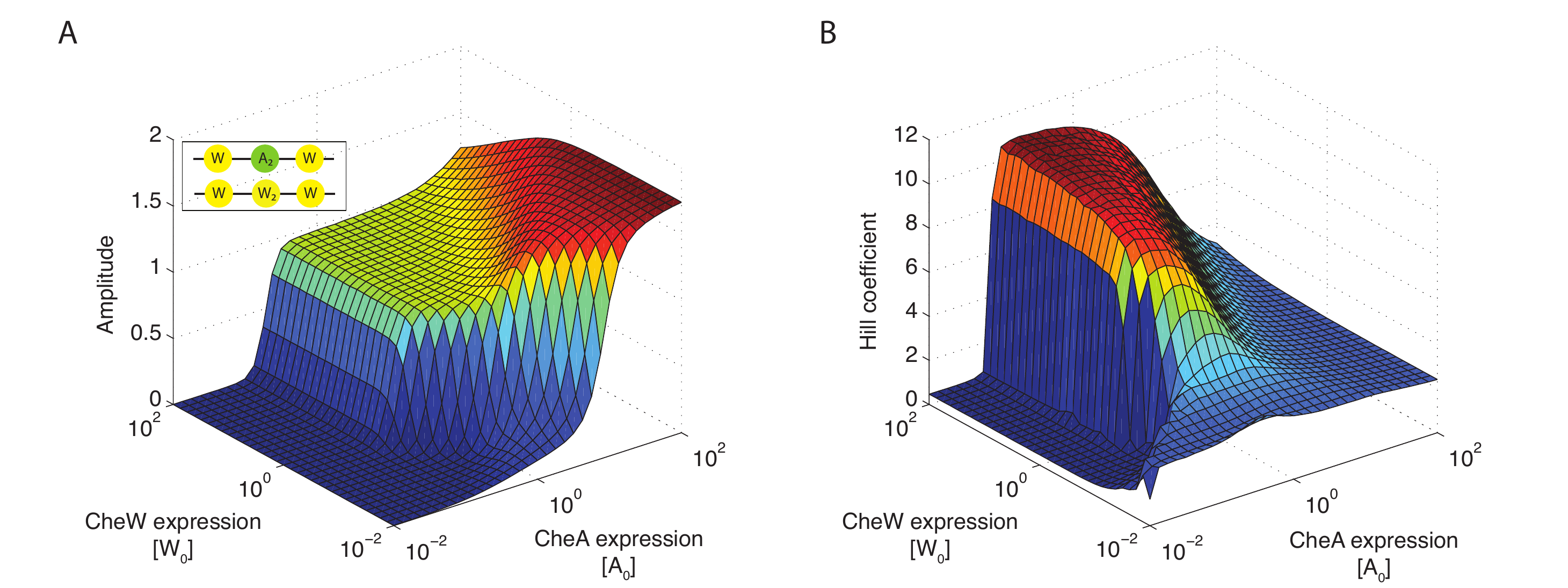}
\caption{
{{\bf Varying CheW and CheA expression levels over a wide range shows trends beyond the FRET data.}  ({\bf A}) Surface plot showing the amplitudes of simulated dose-response curves for different expression levels of CheW and CheA (in units of wild-type levels $\left[W_0\right]$ and $\left[A_0\right]$, respectively).  ({\bf B}) The corresponding surface plot for the Hill coefficient has a saddle-like form. Simulations were performed using the parameter set of Fig. \ref{Fig.ResultsNew} with $\rho= \rho_W$.  }}
\label{Fig.HeatMaps}
\end{figure}

In order to make predictions beyond the data used to fit the model, we created surface plots of amplitudes and Hill coefficients covering several orders of magnitude for expression levels of CheW and CheA (Fig. \ref{Fig.HeatMaps}). 
The receptor activity and hence amplitude increases monotonically with the level of CheA, whereas the increase in amplitude with respect to the level of CheW is only pronounced in a subspace around the experimental data (Fig. \ref{Fig.HeatMaps}A). In the case of high CheA levels, CheW-only linkers exclude CheA from signaling. This also occurs at the wild-type CheA level, although the extent of the effect strongly relies on model parameters. 
The surface plot showing the Hill coefficients as a function of the expression levels of CheW and CheA has a saddle-like form (Fig. \ref{Fig.HeatMaps}B). Although the right flank is consistent with the FRET data at high levels of CheA (small Hill coefficients), the Hill coefficient also decreases at very low levels of CheA as the receptor activity diminishes. 
To test to what extent the model predictions depend on the actual values of parameters $\mu_{\text{W}}^0$, $\mu_{\text{A}_2}^0$ and $\mu_{\text{W}_2}^0$, we varied these parameters and found that the general shape of the surface plot was preserved. Taken together, these observations suggest the need for regulation of both CheW and CheA by the cell to balance signaling amplitude and sensitivity.
Indeed, as CheW and CheA are required in comparable amounts \cite{li_cellular_2004}, both are expressed from the same operon \cite{kollmann_design_2005}.

\section*{Discussion}
Receptor coupling plays a key role in the remarkable sensing and signaling properties of bacterial chemotaxis. These networks can explain the high sensitivity, wide dynamic range and precise adaptation. In this work we present a statistical-mechanics model of different complex sizes, modeling for the first time a molecular linker architecture consistent with (i) FRET dose-response curves, (ii) cryotomography data and (iii) nanodisc experiments. The linker --CheW--CheA\textsubscript{2}--CheW-- proposed by \textit{Li and Hazelbauer} \cite{li_core_2011} is incorporated by assigning expression level-dependent energies $\mu_{\text{W}}$ and $\mu_{\text{A}_2}$ respectively for each CheW and CheA\textsubscript{2} molecule within a complex as part of a fully or partially developed linker. A coupling energy $J<0$ attributed to linkers between active TDs indicates that the coupling between active trimers is stronger than between inactive trimers, in agreement with previous modeling \cite{hansen_dynamic-signaling-team_2010}. Although the actual distribution of complex sizes is influenced by expression levels [W] and [A], a readily adapted chemical potential $\mu$ ensures a fixed average complex size $\rho$ with respect to ligand concentration $c$.  

Our model was first applied to describe the dose-response of Tar receptors in different modification states to MeAsp, a non-metabolizable analog of aspartate. We mainly considered a constant, modification-independent $\rho$, a constraint that not only reduces the number of parameters but also calls into question that the complex size increases with receptor-modification level \cite{endres_variable_2008}. In our work we discovered the discrepancies between the number of connected TDs $N$ and the curves' Hill coefficients $n_\text{H}$ within the classical MWC model. An increase in $N$ is not directly associated with an increase in $n_\text{H}$. In our statistical-mechanics model, the approximately constant $n_\text{H}$ is explained by a constant average complex size across all receptor-modification levels. Indeed, experiments show that both the level of expression of receptors and the insertion of newly synthesized receptors into the inner membrane by the Sec-machinery are highly regulated \cite{gebert_tsr_1988, shiomi_helical_2006}. 

\textit{Hansen et al.} \cite{hansen_dynamic-signaling-team_2010} previously presented a dynamic-signaling-team approach to describe the data obtained with Tar-only cells in which the allosteric coupling among trimers is represented by a modification-dependent trimer-trimer interaction energy $\widehat{J}(m)$ without modeling the actual protein connectivity.
Limited conformational spread and hence a finite complex size is achieved by using a long-range repulsion energy $U$ between all trimers within a complex. In contrast, our model is simpler while providing valuable insights. Neither $\mu_\text{W}$ and $\mu_{\text{A}_2}$ nor $J$ in our ensemble model depend on the modification state of the receptor, and $\mu$ ensures constant average complex size without introducing a repulsive term. Furthermore, the chemical potential $\mu\left(c\right)$ provides insights into the energetic cost of insertion of receptors into the membrane and its dependence on ligand concentration $c$, albeit based on an equilibrium mechanism.

For constant $J$ and $\rho$, we conclude that receptor modification mainly governs the `turn off'-ligand concentration, whereas its influence on receptor clustering is limited. This finding is supported by \textit{Briegel et al.}\cite{briegel_mobility_2013}, who found that the receptor array order and the spacing of receptors in different modification states were indistinguishable. This is in stark contrast to \textit{Hansen et al.} \cite{hansen_dynamic-signaling-team_2010}, who predict a strong increase in average complex size with increasing receptor-modification level. High-resolution imaging of equilibrated receptors in artificial membranes by electron or total internal reflection fluorescence (TIRF) microscopy may allow direct determination of receptor-complex distributions and their dependence on receptor-modification level and ligand concentration. Using photoactivated localization microscopy (PALM) \cite{greenfield_self-organization_2009} or quantitative immunoblotting \cite{li_cellular_2004}, such an investigation could also be performed on intact cells.

Although CheA and CheW have long been known to mediate receptor interactions \cite{sourjik_functional_2004, kentner_determinants_2006}, an increase in   the expression level of CheA leads to a reduction in receptor cooperativity \cite{sourjik_functional_2004}. 
Varying expression levels of CheA and CheW in our model produced results in agreement with experimental data of \textit{Sourjik and Berg} \cite{sourjik_functional_2004}, thereby supporting the linker architecture we employed. The striking observation that increased CheA levels lead to higher kinase activities but lower cooperativity is  based on the fact that the number of CheA dimers per TD is highest for single trimers with almost fully developed linker rest groups (Fig. \ref{Fig.Counting}B). Hence, overexpression of CheA, a bridging molecule at the center of the linker, promotes smaller complex sizes. 
CheA molecules within the rest groups do not contribute to TD coupling and curve steepness, but nevertheless add to the activity of the FRET signal.  

In contrast to what is observed with CheA, raising the level of CheW leads to larger complex sizes and an increased number of empty membrane sites. Again, this behavior becomes comprehensible when the number of CheW molecules per TD (Fig. \ref{Fig.Counting}A) is taken into account. While this ratio is constant for complexes with rest groups, it increases with complex size in the absence of partially developed linkers. Larger complexes directly incorporate more CheA to enhance cooperativity as well as the amplitudes of FRET signals observed both in the model and experimentally. 
In light of our model the experimental observations are produced by a combination of constant receptor density and (partial) linkers. Although partial linkers play a crucial role in the mechanism of our model, their inclusion might appear arbitrary at first. Interestingly, \textit{Briegel et al.}\cite{briegel_new_2014} recently observed a range of assembly intermediates and partial receptor hexagons forming when [W] and [A] were varied. Our surface plots of amplitudes and Hill coefficients also make testable predictions for wide-ranging CheA and CheW expression levels (Fig. \ref{Fig.HeatMaps}).
Is there any evidence to suggest that $\rho$ remains constant when CheA and CheW expression levels change? First, CheA and CheW binding to the receptors occurs after insertion of the receptors into the membrane. Second, increasing the expression of a protein, e.g., of CheW, should remove ribosomes from translating receptor mRNA \cite{scott_emergence_2014,weise_mechanistic_2015}. Although expected to be a minor perturbation, this may lead to a reduced receptor density and hence cooperativity. However, the opposite trend is observed in FRET experiments \cite{sourjik_functional_2004}.

\begin{figure}[!ht]
\includegraphics[width=1.0\textwidth]{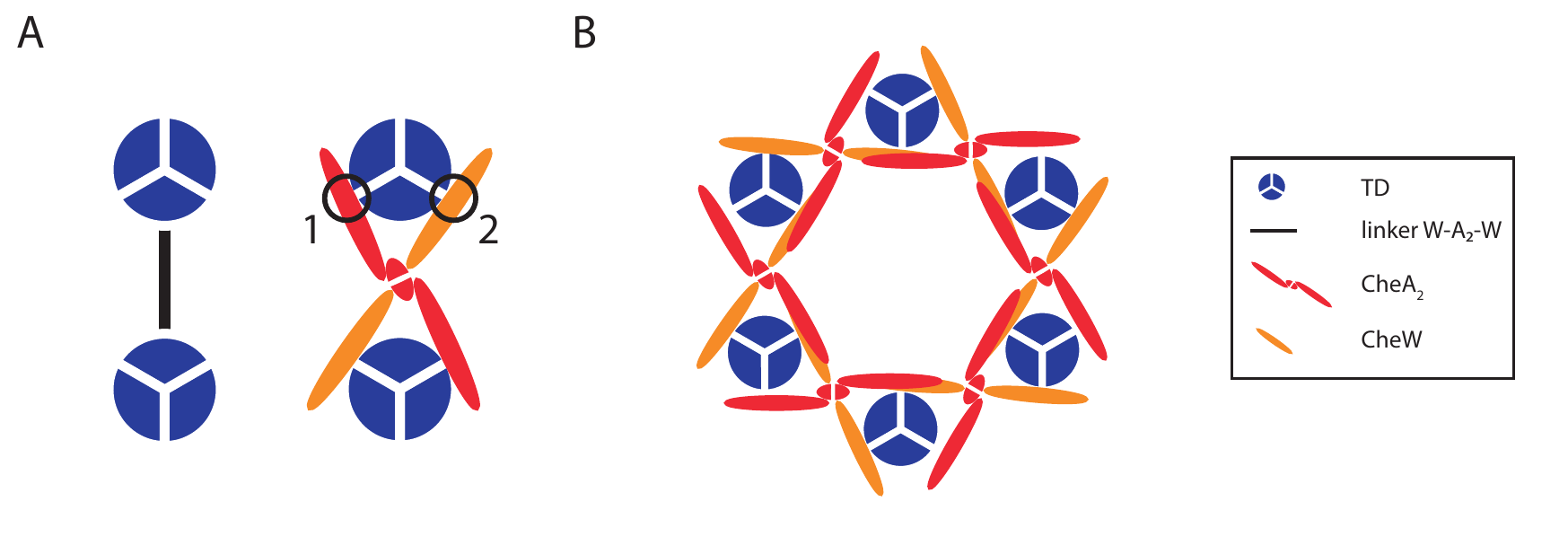}
\caption{
{\bf Structural insights from electron cryotomography.}  ({\bf A}) Our linker --CheW--CheA\textsubscript{2}--CheW-- based on nanodisc experiments (black line) \cite{li_core_2011} appears to be more complicated in reality, where the P5 domain of CheA (homologous to CheW) may also contact the trimers of dimers directly (contact 1 in red) \cite{piasta_increasing_2014, wang_cheareceptor_2012, liu_molecular_2012}. Such an alternative linker, defined as =CheW/CheA\textsubscript{2}/CheW= is explored in \nameref{S4_Fig.}. However, the binding of CheW to the trimers (contact 2 in orange) is presumably much stronger (see Fig. 5A,B in \cite{li_core_2011} and discussion
in \cite{briegel_bacterial_2012}), rendering CheA\textsubscript{2} effectively a bridging molecule. ({\bf B}) Hexagonally packed trimer-of-dimers structure in which the inner connecting ring is formed by alternating CheW/CheA (P5) units \cite{briegel_universal_2009, briegel_bacterial_2012, briegel_new_2014, liu_molecular_2012, briegel_structure_2014}. }
\label{Fig.SchematicsHomology}
\end{figure}

Although our assumed linear linker structure --CheW--CheA\textsubscript{2}--CheW-- matches observed stoichiometries \cite{li_core_2011, li_cellular_2004}, electron cryotomography images suggest that reality is more complicated \cite{briegel_new_2014, briegel_structure_2014}. Modeling of the electron density and spin-labeling studies suggest that CheW and the P5 domain of CheA form alternating CheW/CheA rings connecting the trimers, with P5 occupying positions approximately equivalent to CheW (see Fig. \ref{Fig.SchematicsHomology}). This arrangement is consistent with the strong structural homology between P5 and CheW. However, to describe the FRET data obtained with cells with overexpressed CheA and CheW \cite{sourjik_functional_2004}, our model predicts that CheA\textsubscript{2} has the role of a bridging molecule  and connects trimers via a CheW associated with each trimer. Indeed, an alternative linker with direct receptor-CheA binding and hence symmetric roles of CheA and CheW upon clustering does not match the FRET data (see panel D in \nameref{S4_Fig.}.). 
This view is supported by binding assays, which show that CheW binds much firmer to receptor trimers than CheA to trimers (see Fig. 5A,B in \cite{li_core_2011} and also discussion in \cite{briegel_bacterial_2012}). \\

\begin{figure}[!ht]
\includegraphics[width=1.0\textwidth]{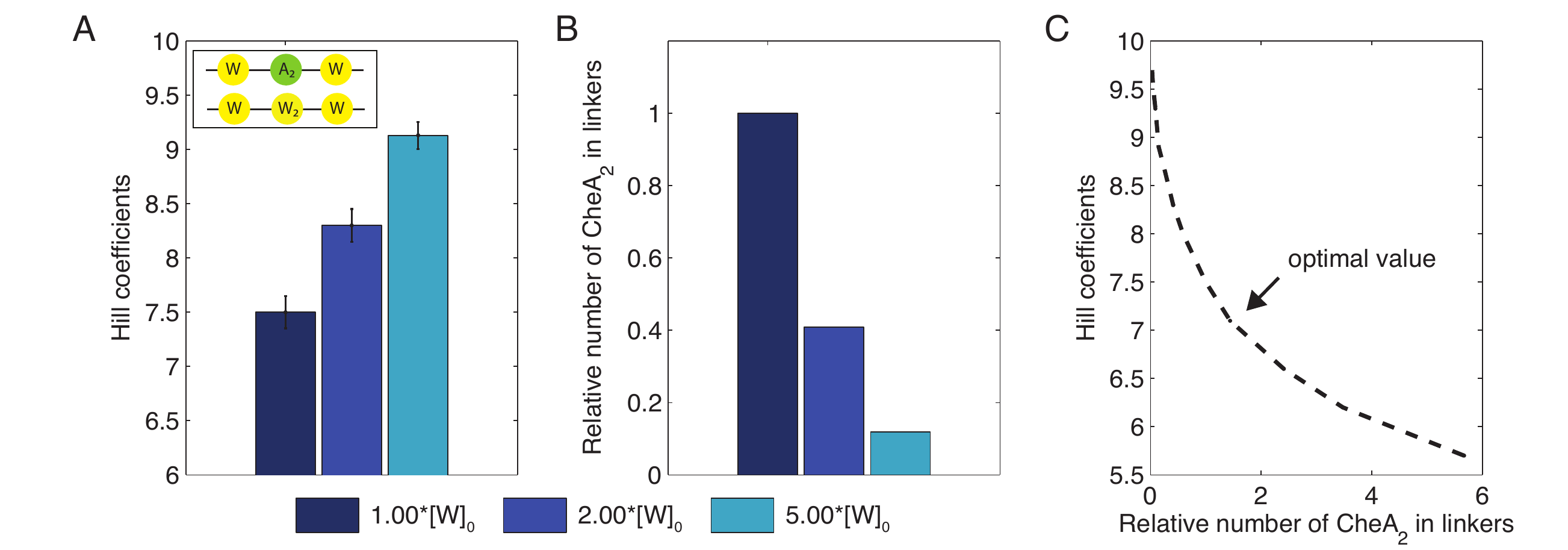}
\caption{
{\bf Simulations suggest that an optimal level of CheW is required for highly cooperative signaling.}  ({\bf A}) Modeled Hill coefficients for different expression levels of CheW as multiples of its native level $[W]_0$. CheA is modeled at its native level $[A]_0$ for all charts. ({\bf B}) The relative number of CheA dimers per linker is simulated for the different levels of CheW expression. Results are compared to the native level $[W]_0$. ({\bf C}) Although increased levels of CheW lead to larger clusters, formation of CheW-only linkers also excludes CheA from signaling. These findings suggest that an optimal CheW level is required to balance signaling sensitivity and magnitude. Model parameters are as in Fig. \ref{Fig.ResultsNew}. }
\label{Fig.Optimal_CheW}
\end{figure}

Although our model qualitatively reproduces the experimental FRET data, the change in cooperativity with variation in [W] is less pronounced in the simulation than in experiments. 
Recent findings based on electron cryotomography offer a possible explanation for this shortcoming. \textit{Briegel et al.}\cite{briegel_new_2014} and \textit{Liu et al.}\cite{liu_molecular_2012} stress the importance of the implemented core unit stoichiometry, but they propose a second type of linker that only involves CheW, with P5/CheW interactions replaced by CheW/CheW interactions \cite{liu_molecular_2012}. To investigate the consequences of these findings for signaling behavior, we allowed for an additional --CheW--CheW\textsubscript{2}--CheW-- linker in our model. 
The simulated dose-response curves show a gre\textit{}atly enhanced change of cooperativity with variation in [W] (Fig. \ref{Fig.ResultsNew}C and Fig. \ref{Fig.Optimal_CheW}A). The generally increased Hill coefficients, and hence sensitivity, may reveal an evolutionary advantage that is not apparent in the tomography images but is detected by FRET. However, whereas CheW-only linkers fit the FRET observations, their incorporation into complexes needs to be tightly regulated. Moreover, in addition to excluding CheA from signaling (Fig. \ref{Fig.Optimal_CheW}B), high levels of CheW were also claimed to disrupt receptor clustering \cite{cardozo_disruption_2010}. Taken together these observations suggest that an optimal level of CheW is required for cooperative signaling by receptors (Fig. \ref{Fig.Optimal_CheW}C). 

In conclusion, our work integrates functional (FRET) and structural (nanodisc and electron cryotomography) data, explains the paradoxes that increased levels of CheA lead to less cooperativity, and provides a functional role for CheW-only linkers. Our proposed linker --CheW--CheA\textsubscript{2}--CheW-- is consistent both with the data from experiments with nanodiscs \cite{li_core_2011} and with images from electron cryotomography \cite{liu_molecular_2012, briegel_bacterial_2012, briegel_new_2014}, if the P5 domain of CheA binds more weakly to the receptor than does CheW. We predict that the observed tetrameric CheW linker, if incorporated at an optimal level, increases the cooperativity while keeping the receptor activity at a sufficiently high level.  An increased understanding of the protein connectivity in receptor clusters may aid not only in describing the fundamental biology of receptor signaling, including the role of cytoplasmic receptor clusters in \textit{Rhodobacter sphaeroides} and \textit{Vibrio cholerae} \cite{briegel_structure_2014}, but may also contribute to the design of novel biosensors \cite{donaldson_development_2014}.

\section*{Materials and Methods}
Keeping $\rho$ constant requires nonlinear optimization of $\mu$ at every ligand concentration. For performance reasons we therefore chose to implement the model in C\# and used a custom-written toolbox to connect to MATLAB 2014a for parameter optimization and plotting. The value for $\mu$ is determined based on Brent's method for root-finding \cite{brent_algorithms_2003}. Fitting of model parameters employs Global Search from MATLAB Global Optimization Toolbox. Multiple start points are generated using scatter-search options (5000 trial points).  For the different start points square deviations from experimental data are minimized using the function \textit{fmincon} with interior point optimization. Note while the number of molecular species in the model increases linearly with the maximal complex size, the computational time is determined by the root finding. The latter becomes considerably harder with additional exponentials of increasing arguments in Eqs.~\ref{eq:Z}, \ref{eq:Ps} and \ref{eq:rho}.

In order to quantify the cooperative behavior of the complexes, Hill functions $A(c)$ (Eq.~\ref{eq:Hill}) with amplitude $A_0$, half-maximum concentration $c_\text{H}$ and Hill coefficient $n_\text{H}$ are fitted to the model evaluated at 50 logarithmically spaced concentrations between $c=0.001$mM and $c=1$mM. 
The Hill coefficients in the comparative plot Fig. \ref{Fig.Tar-only}B result from direct fitting to the experimental data. 
\begin{equation}\label{eq:Hill}
A(c)= \frac{A_0}{1 + \left( \frac{c}{c_\text{H}} \right)^{n_\text{H}}} 
\end{equation}

Though parameter confidence intervals can be calculated based on robust regression and the resulting covariance matrix, especially for highly nonlinear models as ours their validity is questionable given the underlying linear theory \cite{pomerantsev_confidence_1999}. We therefore decided against including confidence intervals except for the fitted Hill curves. 

We note that for all simulations with variations in expression of CheA and CheW the Hill amplitudes match quantitatively much better their experimental counterparts than do the Hill coefficients. 
This observation is partly owed to the fitting routine. With logarithmically spaced concentrations, a difference in amplitude between model and experimental curve directly impacts the corresponding $\chi^2$ goodness-of-fit value. In contrast, a small variation in the Hill coefficient only influences the slope of the curve within a relatively narrow range of ligand concentrations and hence is less reflected in the optimization function value.

\section*{Supporting Information}


\subsection*{S1 Fig. }
\label{S1_Fig.}
\includegraphics[width=1.0\textwidth]{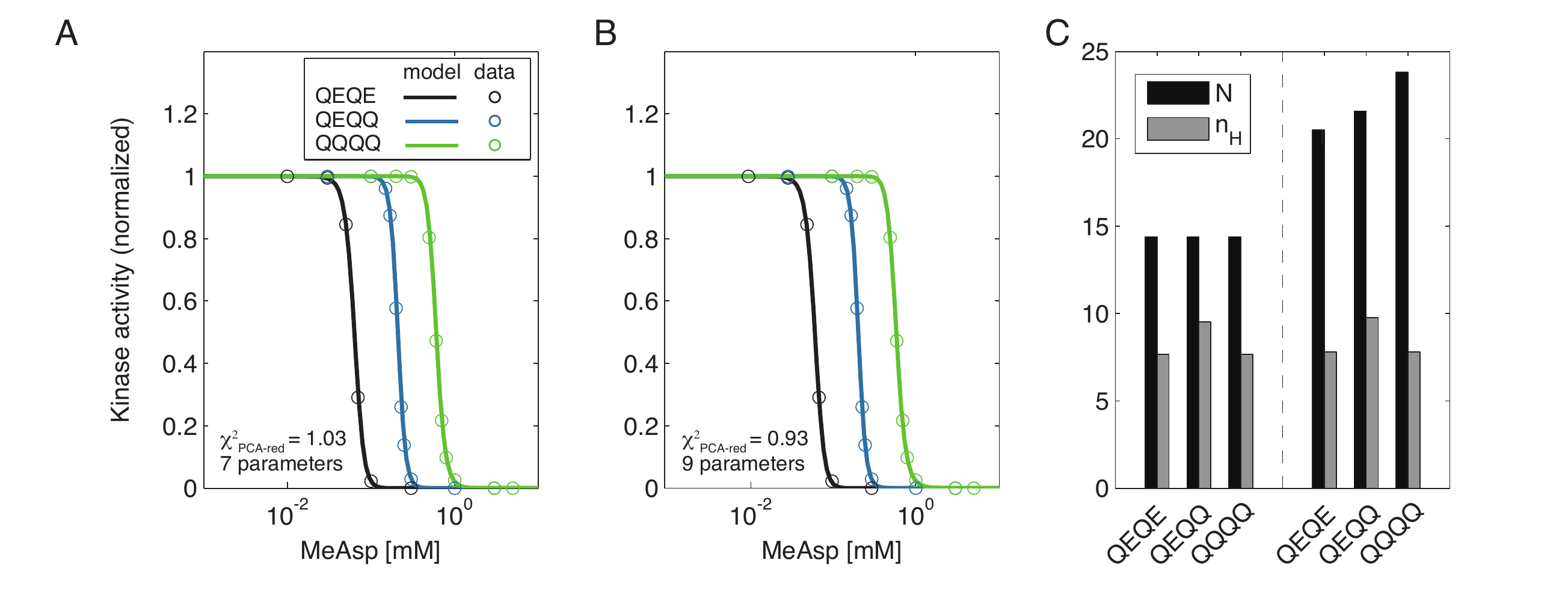} 
{\bf Kinase activity for different Tar-modification levels can be described with a constant receptor-complex size.} ({\bf A,B}) Kinase activity for Tar receptors in QEQE (black), QEQQ (blue) and QQQQ (green) modification states fitted in the classical MWC model with ({\bf A}) constant and ({\bf B}) variable receptor-complex size $N$. The fitting based on Principal Component Analysis (PCA) follows Ref. \cite{endres_variable_2008}. Relating the resulting $\chi^2$ values to the degrees of freedom, here calculated as the number of included PCA components minus the number of model parameters, results in similar goodness-of-fit values $\chi^2_\text{PCA-red}$ with subscript `red' describing the reduced $\chi^2$. However, it should be noted that the actual $\chi^2_\text{PCA-red}$ here is rather a supportive argument to the apparent similarity of both fits, as the number of degrees of freedom is not well defined for nonlinear models \cite{andrae_dos_2010}. 
({\bf C}) Comparison of receptor-complex size $N$ and Hill coefficient $n_H$ for fits with constant $N$ (left, panel A) and variable $N(m)$ (right, panel B).

\subsection*{S2 Fig. }
\label{S2_Fig.}
\includegraphics[width=1.0\textwidth]{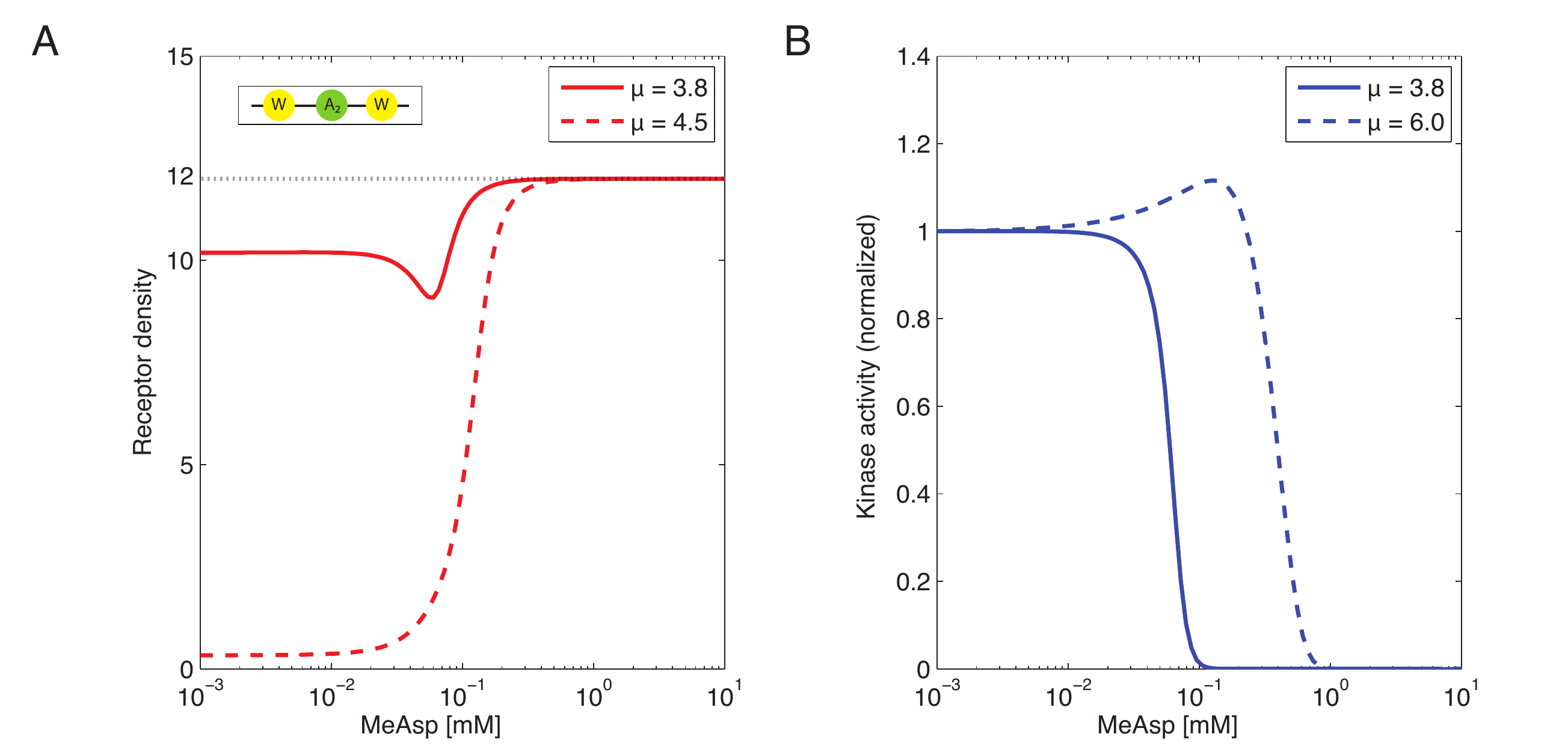} 
{\bf Receptor density increases with ligand concentration for constant chemical potential.} 
({\bf A}) Receptor density as a function of ligand concentration for $\mu=3.8$ (solid) and $\mu=4.5$ (dashed). 
For a constant chemical potential $\mu$, the values of the single dimer energies $f_\text{on}$ and $f_\text{off}$ (Eq.~\ref{eq:fon}) decrease with increasing ligand concentration c. The decrease in the resulting complex energies $F_\text{on}$ and $F_\text{off}$ (Eq.~\ref{eq:Fon}) is stronger for larger complexes. Hence, larger complexes are favored with increasing ligand concentration (Eq.~\ref{eq:Ps}), resulting in an increased receptor density (Eq.~\ref{eq:rho}). 
The interim decrease in $\rho$ for $\mu=3.8$ is the result of an ensemble effect. While the probabilities of all complex sizes increase with c, the increase for larger complexes starts at higher c values. Starting off at a smaller receptor density, this effect is not visible for $\mu=4.5$. Finally both densities asymptotically approach the maximal value of 12. 
({\bf B}) Normalized kinase activity as a function of ligand concentration for $\mu=3.8$ (solid) and $\mu=6.0$ (dashed). In the case of $\mu=3.8$, the increase in receptor density is not apparent in the dose-response curve as the receptors `turn off' before the density increase comes into effect. For $\mu=6.0$, however, the increase in receptor density yields a `bump' in the dose-response curve. 
All plots were generated using the same parameters for QEQE as in Fig. \ref{Fig.Tar-only}.

\subsection*{S3 Fig. }
\label{S3_Fig.}
\includegraphics[width=1.0\textwidth]{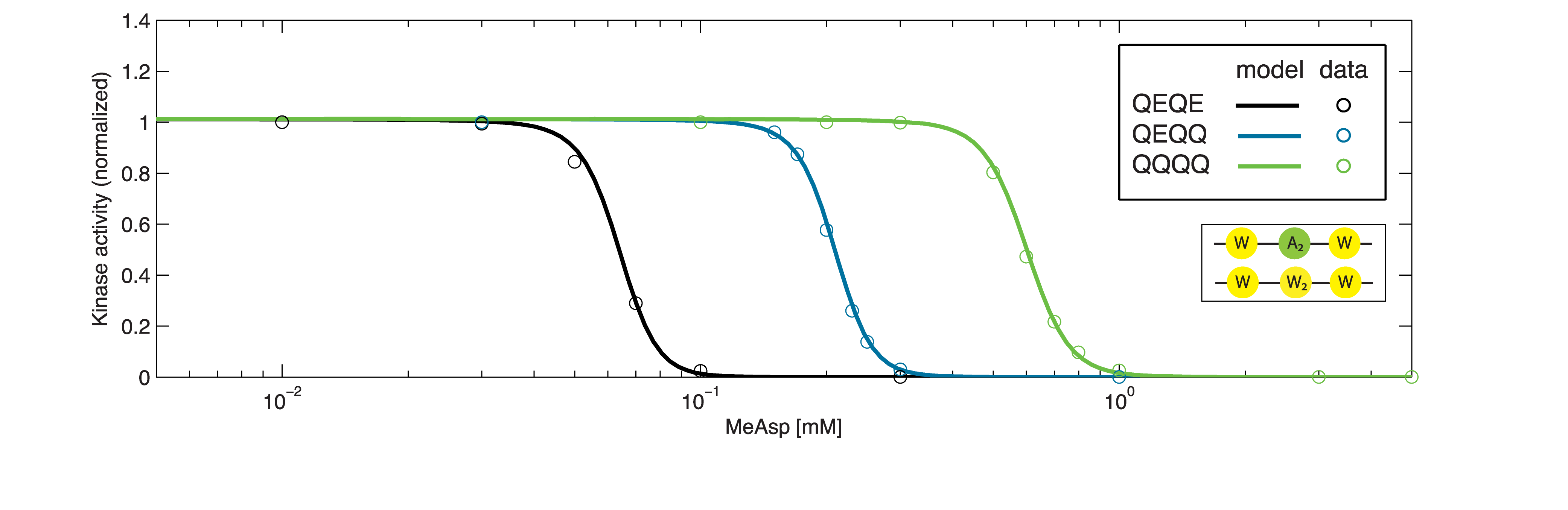} 
{\bf Kinase activity for different Tar-modification levels with additional CheW-only linkers.} Plot following Fig. \ref{Fig.Tar-only}A showing kinase activity for Tar receptors in QEQE (black), QEQQ (blue) and QQQQ (green) modification states. Here the model includes both linkers (--CheW--CheW\textsubscript{2}--CheW-- and --CheW--CheA\textsubscript{2}--CheW--). For simplicity parameters are the same as in 
Fig. \ref{Fig.Tar-only} with the additional value for $\mu_{\text{W}_2}$ in agreement with the value used in Fig. \ref{Fig.ResultsNew}.

\subsection*{S4 Fig. }
\label{S4_Fig.}
\includegraphics[width=1.0\textwidth]{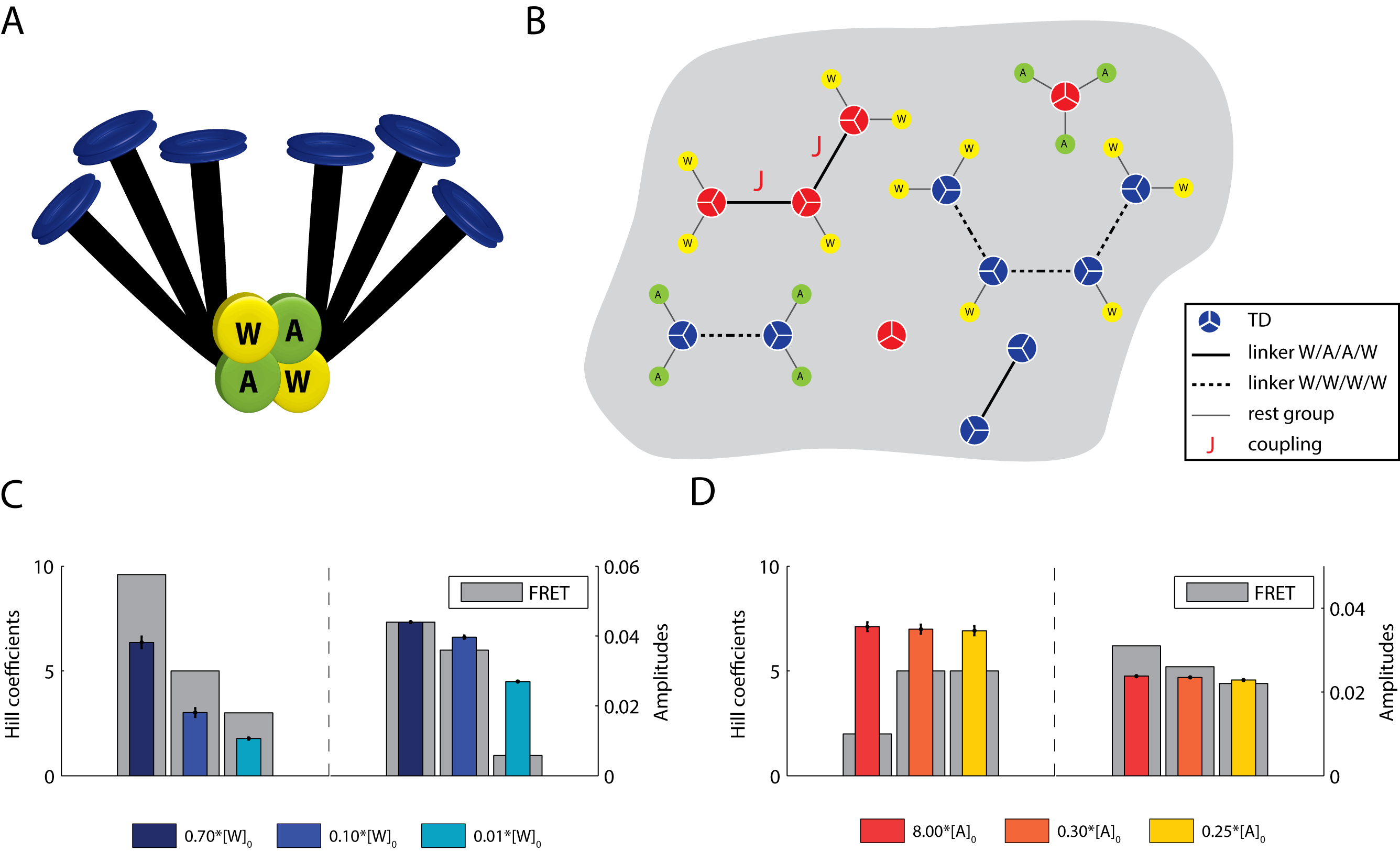} 
{\bf Alternative model with both CheW and CheA binding to trimers does not explain FRET data.}
 ({\bf A}) Schematics of an alternative linker =CheW/CheA\textsubscript{2}/CheW= with both CheA and CheW contacting the trimers directly. ({\bf B}) Exemplary ensemble of complexes in a membrane. The two linkers are represented by solid (=CheW/CheA\textsubscript{2}/CheW=) and dashed (=CheW/CheW\textsubscript{2}/CheW=) lines. Active and inactive TDs are shown in red and blue, respectively. Each linker between active TDs contributes a coupling energy $J$. As monomeric CheA binds directly to trimers, all linker molecule energies are indicated for monomers, hence the linker energy contributions in Eqs. \ref{eq:Fon} become $(x-1)\left(2 \mu_W + 2 \mu_A\right)$ (standard linker) and $(x-1)\left(2 \mu_W + 2 \mu_W\right)$ (CheW-only linker). 
({\bf C,D}) In analogy to Fig. \ref{Fig.ResultsNew}, we fitted the alternative model to the experimental data for varied expression levels of CheW and CheA using a global optimization routine (see \textit{Materials and Methods}). While the alternative model is qualitatively able to describe the effect of changing CheW levels correctly ({\bf C}), it falls short of reproducing the cooperativity decrease for increasing CheA levels ({\bf D}) with nearly identical curves as best fitting result.
Model parameters: $\Delta\epsilon\left(QEQE\right)=-0.23$, $K_\text{D,Tsr}^\text{on}=17.78$, $K_\text{D,Tsr}^\text{off}=0.02$, $\rho_{\text{W}}=2.57$, $\rho_{\text{A}}=3.96$, $\mu_{\text{W}}^0=-1.86$,  $\mu_{\text{A}}^0=-3.82$ and $J=-3.99$.

\section*{Acknowledgments}
The authors thank Moritz Beutel for designing the custom-written toolbox connecting C\# and MATLAB, and Thomas Rogerson and Victor Sourjik for helpful discussions and a critical reading of the manuscript.


%
%
%

\end{document}